\documentclass[english,aps,pra,twocolumn,showpacs]{revtex4-1}
\usepackage{amsmath}
\usepackage{amssymb}
\usepackage{graphicx}
\usepackage{breakurl}
\usepackage[T1]{fontenc}
\usepackage[latin9]{inputenc}
\usepackage{color}
\usepackage{babel}
\usepackage{verbatim}
\usepackage{amstext}
\usepackage{graphicx}
\usepackage[unicode=true,pdfusetitle,pdfborder={0 0 1}]{hyperref}
\usepackage{epstopdf}
\usepackage{bm}
\usepackage{amsfonts}
\usepackage{mathrsfs}

\hypersetup{colorlinks,
linkcolor=blue,          citecolor=blue,        filecolor=blue,      urlcolor=blue           }
\makeatletter
\providecommand{\tabularnewline}{\\}

\@ifundefined{textcolor}{}
{
 \definecolor{BLACK}{gray}{0}
 \definecolor{WHITE}{gray}{1}
 \definecolor{RED}{rgb}{1,0,0}
 \definecolor{GREEN}{rgb}{0,1,0}
 \definecolor{BLUE}{rgb}{0,0,1}
 \definecolor{CYAN}{cmyk}{1,0,0,0}
 \definecolor{MAGENTA}{cmyk}{0,1,0,0}
 \definecolor{YELLOW}{cmyk}{0,0,1,0}
}
\providecommand{\tabularnewline}{\\}
\@ifundefined{textcolor}{}{ \definecolor{BLACK}{gray}{0}
 \definecolor{WHITE}{gray}{1}
 \definecolor{RED}{rgb}{1,0,0}
 \definecolor{GREEN}{rgb}{0,1,0}
 \definecolor{BLUE}{rgb}{0,0,1}
 \definecolor{CYAN}{cmyk}{1,0,0,0}
 \definecolor{MAGENTA}{cmyk}{0,1,0,0}
 \definecolor{YELLOW}{cmyk}{0,0,1,0}
}
\makeatother
\begin{document}

\title{Direct Probe of Topological Order for Cold Atoms}

\author{Dong-Ling Deng}
\affiliation{Department of Physics, University of Michigan, Ann Arbor, Michigan 48109, USA}
\affiliation{Center for Quantum Information, IIIS, Tsinghua University, Beijing 100084,
PR China}
\email{dldeng@umich.edu}

\author{Sheng-Tao Wang}
\affiliation{Department of Physics, University of Michigan, Ann Arbor, Michigan 48109, USA}
\affiliation{Center for Quantum Information, IIIS, Tsinghua University, Beijing 100084,
PR China}

\author{L.-M. Duan}
\affiliation{Department of Physics, University of Michigan, Ann Arbor, Michigan 48109, USA}
\affiliation{Center for Quantum Information, IIIS, Tsinghua University, Beijing 100084,
PR China}

\date{\today }
\begin{abstract}
Cold-atom experiments in optical lattices offer a versatile platform
to realize various topological quantum phases. A key challenge in
those experiments is to unambiguously probe the topological order.
We propose a method to directly measure the characteristic topological
invariants (order) based on the time-of-flight imaging of cold atoms.
The method is generally applicable to detection of topological band
insulators in one, two, or three dimensions characterized by integer
topological invariants. Using detection of the Chern number for the
2D anomalous quantum Hall states and the Chern-Simons term for the
3D chiral topological insulators as examples, we show that the proposed
detection method is practical, robust to typical experimental imperfections
such as limited imaging resolution, inhomogeneous trapping potential,
and disorder in the system.
\end{abstract}


\maketitle
The study of topological phases of matter, such as topological band
insulators and superconductors, has attracted a lot of interest in
recent years \cite{hasan2010colloquium,qi2011topological,moore2010birth}.
Various topological phases have been found associated with the free-fermion
band theory and classified into a periodic table according to the
system symmetry and dimensionality \cite{schnyder2008classification,kitaev2009periodic,ryu2010topological}.
The topology of the band structure is characterized by a topological
invariant taking only integer values, which gives the most direct
and unambiguous signal of the corresponding topological order. To
experimentally probe the topological order, it is desirable to have
a way to measure the underlying topological invariant. For some phase,
the topological invariant may manifest itself through certain quantized
transport property or characteristic edge state behavior \cite{thouless1982quantized}.
For instance, the quantized Hall conductivity is proportional to the
underlying topological Chern number that characterizes the integer
quantum Hall states \cite{thouless1982quantized,von1986quantized,sarma2008perspectives}.
For many other topological phases in the periodic table, it is not
clear yet how to experimentally extract information of the underlying
topological invariants.

Cold atoms in optical lattices provide a powerful experimental platform
to simulate various quantum states of matter. In particular, recent
experimental advance in engineering of spin-orbit coupling and artificial
gauge field for cold atoms \cite{lin2009synthetic,liu2009effect,wang2012spin,cheuk2012spin,dalibard2011colloquium,galitski2013spin}
has pushed this system to the forefront for realization of various
topological quantum phases \cite{lewenstein2007ultracold,bloch2012quantum,zhu2006spin,zhu2011probing,beri2011z_,aidelsburger2013realization,miyake2013realizing}.
The detection method for cold-atom experiments is usually quite different
from those for conventional solid-state materials. A number of intriguing
proposals have been made for detection of certain topological order
in cold-atom experiments, such as those based on the dynamic response
\cite{shao2008realizing,dauphin2013extracting,Hauke2014Tomography,Wang2013ChargeP}, the Bragg spectroscopy
\cite{liu2010quantum,goldman2012detecting}, imaging of the edge states
\cite{goldman2013direct}, counting peaks in the momentum distribution \cite{Zhao2011Chern} or detection of the Berry phase or curvature
\cite{alba2011seeing,price2012mapping,liu2013detecting,abanin2013interferometric,Atala2013direct,Goldman2013Measuring,Hauke2014Tomography,Pachos2013,Lisle2014}.
Most of these proposals are targeted to detection of the quantum Hall
phase. Similar to solid-state systems, it is not clear yet how to
probe the topological invariants for various other topological phases
in the periodic table.

In this Rapid Communication, we propose a general method to directly measure the
topological invariants in cold-atom experiments based on the state-of-the-art
time-of-flight (TOF) imaging. The TOF imaging, combined with the quench dynamics from
the Hamiltonian, has been exploited in recent schemes for detection of the Chern numbers
associated with two-band topological models in one or two dimensional optical lattices \cite{Hauke2014Tomography}.
Compared with the previous work, our method has the following distinctive features:
1) it is applicable to detection of any topological band insulators with spin degrees of freedom
in one, two, or three dimensions that are characterized by integer
topological invariants in the periodic table. 2) The method is not
limited by the requirement of a two-band structure for the Hamiltonian
\cite{Hauke2014Tomography,alba2011seeing} or occupation of only the lowest band \cite{liu2013detecting}.
Instead, it detects the topological invariants associated with each
band for any multi-band Hamiltonians. 3) Our proposed detection method
is very robust to practical experimental imperfections. As examples,
we numerically simulate two experimental detections: one for the Chern
number of the 2D anomalous quantum Hall phase and the other for the
Chern-Simons term of the 3D chiral topological insulator. Both simulations
show that accurate values of the topological invariants can be obtained
experimentally under imaging resolution of a few to a dozen pixels
along each spatial dimension, even with inhomogeneous traps and random
potentials or interactions. The robustness is also found in Ref. \cite{Hauke2014Tomography}
for detection of the Chern number in a different 2D model using the tomography method.

The topological band insulators are described by effective free-fermion
Hamiltonians, typically with complicated spin-orbit couplings. We
consider a real-space Hamiltonian with $N$ spin (pseudo-spin) degrees of freedom,
referred as $|m\rangle$ with $m=1,2,...,N$. In the momentum $\mathbf{k}$ space, the Hamiltonian
has $N$ bands and is described by an $N$-by-$N$ Hermitian matrix
$H(\mathbf{k})$. The energy spectrum is obtained by solving the Schr$\ddot{\text{o}}$inger
equation in the momentum space
\begin{equation}
H(\mathbf{k})|u_{b}(\mathbf{k})\rangle=E_{b}(\mathbf{k})|u_{b}(\mathbf{k})\rangle,\label{1}
\end{equation}
where $b=1,2,\cdots,N$ is the band index and $|u_{b}(\mathbf{k})\rangle$
denotes the corresponding Bloch state with eigen-energy $E_{b}(\mathbf{k})$.
For simplicity, we assume the bands are non-degenerate. Expressed
in the original spin basis $|m\rangle$, the Bloch state has the form
\begin{equation}
|u_{b}(\mathbf{k})\rangle=\sum_{m=1}^{N}c_{bm}(\mathbf{k})|m\rangle,\label{2}
\end{equation}
where $c_{bm}(\mathbf{k})$ is the Bloch wavefunction with normalization
$\sum_{m}|c_{bm}(\mathbf{k})|^{2}=1$.

An topological invariant can be defined for each band, which usually
takes the form of the Chern numbers for even spatial dimensions and
the Chern-Simons terms (or the winding numbers in certain cases) for
odd spatial dimensions. The Chern numbers (or Chern-Simons terms)
can be expressed as momentum-space integrals of the Berry curvature
and connection associated with the Bloch state $|u_{b}(\mathbf{k})\rangle$.
For instance, in 2D ($x,y$-plane), the Chern number $C_{b}$ for
the band $b$ is defined by
\begin{equation}
C_{b}=-\frac{1}{2\pi}\int_{\text{BZ}}dk_{x}dk_{y}F_{xy}^{(b)}(\mathbf{k}),\label{3}
\end{equation}
where the Berry curvature $F_{xy}^{(b)}(\mathbf{k})\equiv\partial_{k_{x}}A_{y}^{(b)}(\mathbf{k})-\partial_{k_{y}}A_{x}^{(b)}(\mathbf{k})$
and the Berry connection $A_{\mu}^{(b)}(\mathbf{k})\equiv\langle u_{b}(\mathbf{k})|i\partial_{k_{\mu}}|u_{b}(\mathbf{k})\rangle$
$(\mu=x,y)$, and the integration is over the whole Brillouin zone
(BZ) which forms a compact manifold.

To probe the topological invariant such as the Chern number in Eq.
(3), what we need to measure is the Bloch wave function $c_{bm}(\mathbf{k})$.
The Berry connection and curvature can be obtained as derivatives
of $c_{bm}(\mathbf{k})$ and the Chern number is just a two-fold integration
of $F_{xy}^{(b)}(\mathbf{k})$. For cold atoms in an optical lattice,
we can map out the momentum distribution with the conventional time-of-flight
imaging and separate different spin components through a magnetic
field gradient \cite{ketterle2008making}. Through the band mapping
technique employed in experiments \cite{bloch2012quantum}, populations
in different bands are mapped to atomic densities in different spatial
regions, so by this measurement we can get information about $n_{bm}(\mathbf{k})=\left\vert c_{bm}(\mathbf{k})\right\vert ^{2}$
for all occupied bands. To extract the wavefunction $c_{bm}(\mathbf{k})$,
it is also crucial to measure the phase information. For this purpose,
we apply an impulsive pulse right before the flight of atoms to induce
a rotation between different spin components \cite{duan2006detecting}.
The rotation should keep the atomic momentum unchanged but mix their
spins. For instance, a $\pi/2$-rotation between spin components $m$
and $m^{\prime}$ induces the transition $c_{bm}(\mathbf{k})\rightarrow\left[c_{bm}(\mathbf{k})+c_{bm^{\prime}}(\mathbf{k})\right]/\sqrt{2}$
and $c_{bm^{\prime}}(\mathbf{k})\rightarrow\left[-c_{bm}(\mathbf{k})+c_{bm^{\prime}}(\mathbf{k})\right]/\sqrt{2}$,
which can be achieved by applying two co-propagating Raman beams or
a radio frequency pulse that couples the spin components $m,m^{\prime}$
and preserves the momentum $\mathbf{k}$. The pulse is short so that
expansion of the atomic cloud is negligible during the pulse. For Raman pairs, relative phase coherence is kept but absolute phase locking is not necessary. The angular momentum change associated with the spin flip can be transferred from the Raman pair and selection rules have to be followed according to the specific atomic levels used.  With
this prior $\pi/2$-pulse, the time-of-flight (TOF) imaging then measures
the densities $\left\vert c_{bm}(\mathbf{k})\pm c_{bm^{\prime}}(\mathbf{k})\right\vert ^{2}/2$,
whose difference gives the interference terms $\texttt{Re}\left[c_{bm}^{\ast}(\mathbf{k})c_{bm^{\prime}}(\mathbf{k})\right]$.
By the same method but with a different phase of the $\pi/2$-pulse,
one can similarly measure the imaginary part $\texttt{Im}\left[c_{bm}^{\ast}(\mathbf{k})c_{bm^{\prime}}(\mathbf{k})\right]$
between any two spin components $m$ and $m^{\prime}$. The measurement
of the population and interference terms $c_{bm}^{\ast}(\mathbf{k})c_{bm^{\prime}}(\mathbf{k})$
for all $m,m^{\prime}$ fully determines the Bloch wave function $c_{bm}(\mathbf{k})$
up to arbitrariness of an overall phase $c_{bm}(\mathbf{k})\rightarrow c_{bm}(\mathbf{k})e^{i\varphi(\mathbf{k})}$,
where $\varphi(\mathbf{k})$ in general is $\mathbf{k}$-dependent
but independent of the spin index.

In experiments, one needs to discretize the TOF image and measure
the density distribution at each pixel of the BZ. The wavefunction
$c_{bm}(\mathbf{k})$ is fixed up to an overall phase $\varphi(\mathbf{k})$
at each pixel with the above method. This arbitrary $\mathbf{k}$-dependent
phase poses an obstacle to measurement of the topological invariants.
To overcome this difficulty, we use a different way to calculate the
Berry curvature based on the so-called $U(1)$-link defined for each
pixel $\mathbf{k}_{\mathbf{J}}$ of the discrete BZ \cite{fukui2005chern}.
The $U(1)$-link is defined as $U_{\nu}^{(b)}(\mathbf{k}_{\mathbf{J}})\equiv\langle u_{b}(\mathbf{k}_{\mathbf{J}})|u_{b}(\mathbf{k_{\mathbf{J+\hat{\nu}}}})\rangle/|\langle u_{b}(\mathbf{k}_{\mathbf{J}})|u_{b}(\mathbf{k_{\mathbf{J+\hat{\nu}}}})\rangle|$,
where $\hat{\nu}=\hat{x},\hat{y},\hat{z}$, a unit vector in the corresponding
direction. A gauge-independent field is obtained from the $U(1)$-link
as \cite{fukui2005chern}
\begin{eqnarray}
\mathcal{F}_{\mu\nu}^{(b)}(\mathbf{k}_{\mathbf{J}}) & \equiv & i\ln\dfrac{U_{\mu}^{(b)}(\mathbf{k}_{\mathbf{J}})U_{\nu}^{(b)}(\mathbf{k}_{\mathbf{J+\hat{\mu}}})}{U_{\mu}^{(b)}(\mathbf{k}_{\mathbf{J+\hat{\nu}}})U_{\nu}^{(b)}(\mathbf{k}_{\mathbf{J}})},\label{4}
\end{eqnarray}
where $\mathcal{F}_{\mu\nu}^{(b)}(\mathbf{k}_{\mathbf{J}})\in(-\pi,\pi]$
corresponds to a discrete version of the Berry curvature and it reduces
to the latter in the large size limit. $\mathcal{F}_{\mu\nu}^{(b)}(\mathbf{k}_{\mathbf{J}})$
can be obtained directly from the TOF images associated with the pixel
$\mathbf{k}_{\mathbf{J}}$ of the BZ, independent of the overall phase
factor $\varphi(\mathbf{k})$. The topological invariant can be calculated
from $\mathcal{F}_{\mu\nu}^{(b)}(\mathbf{k}_{\mathbf{J}})$ by a direct
summation over all the pixels of the BZ (instead of $\mathbf{k}$
integration in Eq. (3)). This gives a simple and robust way to experimentally
extract the topological invariant from the TOF images.

The detection method described above is general and applicable to
various topological phases in different spatial dimensions. To show
that the method is robust to experimental imperfections, in the following
we numerically simulate detection of two kinds of topological invariants:
one is the Chern number associated with the 2D quantum anomalous Hall
effect and the other is the Chern-Simons term associated with the
3D chiral topological insulator.

\begin{figure}
\includegraphics[width=0.49\textwidth]{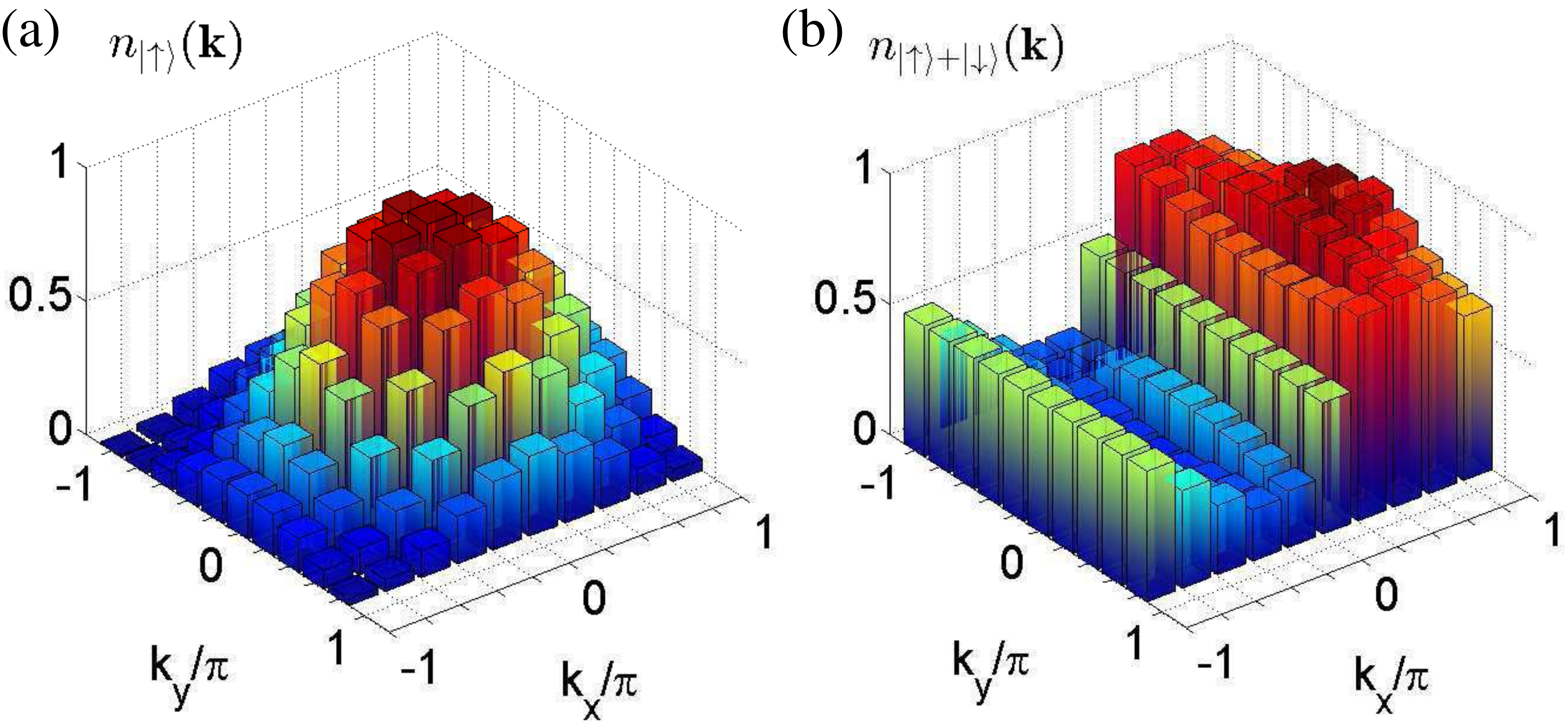} \caption{(color online). Density distributions in momentum space for the first
band of $H_{\text{QAH}}$ under two different spin bases with lattice
size $10\times10$ and open boundary condition. The total density
at each $\mathbf{k}$ is normalized to unity (e.g.,\ $n_{|\uparrow\rangle}(\mathbf{k})+n_{|\downarrow\rangle}(\mathbf{k})=1$)
corresponding to the unit filling. The parameters are chosen to be
$\lambda_{\text{SO}}^{(x)}=\lambda_{\text{SO}}^{(y)}=h=t$, $\gamma_{T}=0.01t$,
and $\gamma_{P}=0.1t$.}

\label{fig: QAH-DensityDis}
\end{figure}

\textit{2D quantum anomalous Hall (QAH) effect---} The conventional
quantum Hall effect requires application of a strong magnetic field.
For the QAH effect, a combination of spontaneous magnetization and
spin-orbit coupling gives rise to quantized Hall conductivity in the
absence of an external magnetic field \cite{haldane1988model}. In solid-state systems, a recent
experiment has observed this peculiar phenomenon in thin films of a
magnetically doped topological insulator \cite{chang2013experimental}. A
simple square-lattice Hamiltonian which captures the essential physics
of the QAH effect has the following form in real space:
\begin{align}
H_{\text{QAH}}  &=  \lambda_{\text{SO}}^{(x)}\sum_{\mathbf{r}}[(a_{\mathbf{r}\uparrow}^{\dagger}a_{\mathbf{r}+\hat{x}\downarrow}-a_{\mathbf{r}\uparrow}^{\dagger}a_{\mathbf{r}-\hat{x}\downarrow})+\text{H.c.}]
\label{5}\\
&+ i\lambda_{\text{SO}}^{(y)}\sum_{\mathbf{r}}[(a_{\mathbf{r}\uparrow}^{\dagger}a_{\mathbf{r}+\hat{y}\downarrow}-a_{\mathbf{r}\uparrow}^{\dagger}a_{\mathbf{r}-\hat{y}\downarrow})+\text{H.c.}]\notag\\
 &- t\sum_{\langle\mathbf{r},\mathbf{s}\rangle}(a_{\mathbf{r}\uparrow}^{\dagger}a_{\mathbf{s}\uparrow}-a_{\mathbf{r}\downarrow}^{\dagger}a_{\mathbf{s}\downarrow})+h\sum_{\mathbf{r}}(a_{\mathbf{r}\uparrow}^{\dagger}a_{\mathbf{r}\uparrow}-a_{\mathbf{r}\downarrow}^{\dagger}a_{\mathbf{r}\downarrow}),\notag
\end{align}
where $a_{\mathbf{r}\sigma}^{\dagger}$$(a_{\mathbf{r}\sigma})$ is
the creation (annihilation) operator of the fermionic atom with pseudospin
$\sigma=(\uparrow,\downarrow)$ at site $\mathbf{r}$, and $\hat{x},\hat{y}$
are unit lattice vectors along the $x,y$ directions. The first term
in the Hamiltonian describes the spin-orbit coupling. The second and
the third terms denote respectively the spin-conserved nearest-neighbor
hopping and the Zeeman interaction. It was proposed recently that
$H_{\text{QAH}}$ may be realized with cold fermionic atoms trapped
in a blue-detuned optical lattice \cite{Liu2014Realization}.

In momentum space, this Hamiltonian has two Bloch bands. The topological
structure of this model is characterized by the Chern number defined
in Eq. (3). Direct calculation shows that $C_{2}=-C_{1}=\text{sign}(h)$
when $0<|h|<4t$ and $C_{2}=-C_{1}=0$ otherwise. Experimentally,
one can measure $\mathcal{F}_{\mu\nu}^{(b)}(\mathbf{k}_{\mathbf{J}})$
by our proposed method to extract the Chern number through $C_{b}\approx-\sum_{\mathbf{J}}\mathcal{F}_{xy}^{(b)}(\mathbf{k}_{\mathbf{J}})/\left(2\pi\right)$,
where the band index $b=1,2$.

To simulate experiments, we consider a finite-size lattice with open
boundary condition. In addition, we add a global harmonic trap of
the form $V_{T}=m_{a}\omega^{2}r^{2}/2$ for atoms of mass $m_{a}$
as in real experiments and use $\gamma_{T}=m_{a}\omega^{2}a^{2}/2t$
to parameterize the relative strength of the trap, where $a$ denotes
the lattice constant. To account for possible experimental noise,
we add a random perturbation Hamiltonian of the following general
form
\begin{equation}
H_{\text{P}}=\gamma_{\text{P}}t\sum_{\left\langle \mathbf{r},\mathbf{s}\right\rangle ,\alpha,\beta}a_{\mathbf{r},\alpha}^{\dagger}\mathcal{P}_{\mathbf{r\alpha},\mathbf{s}\beta}a_{\mathbf{s},\beta},\label{6}
\end{equation}
where $\gamma_{\text{P}}$ is a dimensionless parameter characterizing
the strength of random perturbation, $\left\langle \mathbf{r},\mathbf{s}\right\rangle $
denotes the neighboring sites, and $\mathcal{P}$ is a random Hermitian
matrix with its largest eigenvalue normalized to unity. We numerically
diagonalize the real-space Hamiltonian on a finite lattice with different
number of sites and calculate the corresponding momentum density distributions
\cite{supplement}. An an example, in Fig.\ \ref{fig: QAH-DensityDis},
we show the reconstructed density distribution in two complementary
bases ($\left\{ \left\vert \uparrow\right\rangle ,\left\vert \downarrow\right\rangle \right\} $,
$\left\{ \left\vert \uparrow\right\rangle \pm\left\vert \downarrow\right\rangle \right\} $)
under open boundary condition with a harmonic trap and random perturbations
(more detailed calculation results are shown in the the supplement
\cite{supplement}). The Chern numbers for each case are calculated
and listed in Table \ref{Numerics-Table} under choices of different
parameters and system sizes. The extracted Chern numbers exactly equal
the corresponding theoretical values, even under a small system size
and significant disorder potentials. This is so as Chern numbers characterize
the topological property, which does not change under perturbations.
Furthermore, our detection method through measurement of $\mathcal{F}_{\mu\nu}^{(b)}(\mathbf{k}_{\mathbf{J}})$
guarantees an integer value for the extracted Chern number \cite{fukui2005chern},
so it automatically corrects small errors due to experimental imperfections. Ref.\ \cite{Hauke2014Tomography} also points out the robustness of Fukui \emph{et al}.'s method \cite{fukui2005chern} in computing the Chern number.

\begin{table}
\caption{Simulated detection results of the topological invariants for different
lattice sizes under various conditions (Periodic boundary condition,
Open boundary condition, with Trap, with both Trap and Perturbation
Hamiltonians). For the QAH, the invariant is the Chern number for
the first band ($C_{1}$), whereas for the CTI, it is the Chern-Simons
term for the middle flat band ($CS_{2}/\pi$). Results for both the
nontrivial phase ($h/t=1$ for the QAH and $h/t=2$ for the CTI) and
the trivial phase ($h/t=5$ for the QAH and $h/t=4$ for the CTI)
are presented. The parameters are the same as in Fig.\ \protect\ref{fig: QAH-DensityDis}
and Fig.\ \protect\ref{fig:CTI-Density Distr}. }

\label{Numerics-Table}\begin{ruledtabular}%
\begin{tabular}{ccccccc}
 & h/t  & Size  & Periodic  & Open  & Trap  & Pert.+Trap \tabularnewline
\hline
QAH  & 1  & $4^{2}$  & -1  & -1  & -1  & -1 \tabularnewline
 & 1  & $10^{2}$  & -1  & -1  & -1  & -1 \tabularnewline
 & 5  & $10^{2}$  & 0  & 0  & 0  & 0 \tabularnewline
 &  &  &  &  &  & \tabularnewline
CTI  & 2  & $10^{3}$  & $1.041$  & $1.056$  & $1.055$  & $1.080$ \tabularnewline
 & 2  & $12^{3}$  & $1.031$  & $1.009$  & $0.981$  & $1.014$ \tabularnewline
 & 4  & $10^{3}$  & $0$  & $-2*10^{-4}$  & $1.1*10^{-3}$  & $1.2*10^{-3}$\tabularnewline
\end{tabular}\end{ruledtabular}
\end{table}

\textit{3D chiral topological insulator---} Chiral topological insulators
(CTIs) are protected by the chiral symmetry (also known as the sub-lattice
symmetry) and belong to the AIII class in the periodic table for topological
phases \cite{kitaev2009periodic,schnyder2008classification,ryu2010topological}.
A simple Hamiltonian that supports 3D CTIs has the form \cite{neupert2012noncommutative}:
\begin{eqnarray}
H_{\text{CTI}} & = & \frac{t}{2}\sum_{\mathbf{r}}\sum_{j=1}^{3}[\psi_{\mathbf{r}}^{\dagger}(iG_{3+j}-G_{7})\psi_{\mathbf{r}+\mathbf{e}_{j}}+\text{H.c.}]\notag\\
 &  & +h\sum_{\mathbf{r}}\psi_{\mathbf{r}}^{\dagger}G_{7}\psi_{\mathbf{r}},\label{7}
\end{eqnarray}
where the operator $\psi_{\mathbf{r}}^{\dagger}=(a_{\mathbf{r},1}^{\dagger},a_{\mathbf{r},2}^{\dagger},a_{\mathbf{r},3}^{\dagger})$
with $a_{\mathbf{r},\alpha}^{\dagger}$$(\alpha=1,2,3)$ creating
a fermion at site $\mathbf{r}$ with spin state $\alpha$, $\mathbf{e}_{1},\mathbf{e}_{2},\mathbf{e}_{3}$
are unit vectors along the $x,y,z$ directions, and $G_{\nu}$ ($\nu=4,5,6,7$)
denotes the $\nu$th Gell-Mann matrix \cite{supplement}. In the momentum
space, this model Hamiltonian has three gapped bands, with a zero-energy
flat band in the middle protected by the chiral symmetry. An experimental
scheme has been proposed to realize this model Hamiltonian with cold
fermionic atoms in an optical lattice \cite{Wang2014probe}. The topological
property of this Hamiltonian can be described by the Chern-Simons
term. For the $b$-th ($b=1,2,3)$ Bloch band, the Chern-Simons term
$CS_{b}$ takes the form
\begin{equation}
CS_{b}=-\frac{1}{4\pi}\int_{\text{BZ}}d\mathbf{k}\epsilon^{\mu\nu\tau}A_{\mu}^{(b)}(\mathbf{k})\partial_{k_{\nu}}A_{\tau}^{(b)}(\mathbf{k}),\label{8}
\end{equation}
where $A_{\mu}^{(b)}(\mathbf{k})=\langle u_{b}(\mathbf{k})|\partial_{k_{\mu}}|u_{b}(\mathbf{k})\rangle$
$(\mu=x,y,z).$ Explicit calculations show that $CS_{3}=CS_{1}=CS_{2}/4=\pi\Gamma(h)/4$
with $\Gamma(h)=-2$ for $|h|<t,$ $\Gamma(h)=1$ for $t<|h|<3t$,
and $\Gamma(h)=0$ otherwise.

\begin{figure}
\includegraphics[width=0.49\textwidth]{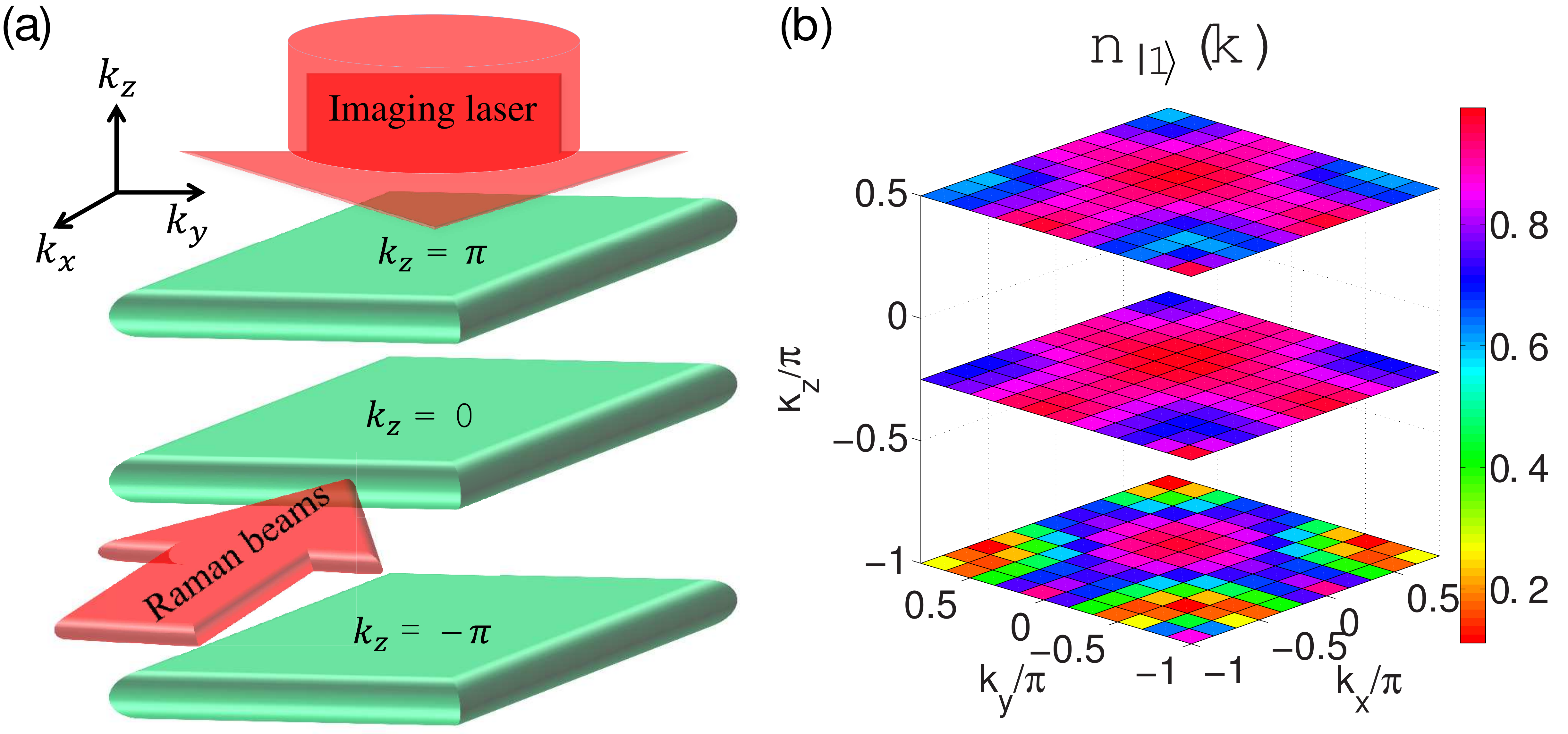} \caption{(color online). (a) An illustration to show reconstruction of the
3D atomic momentum distribution by the TOF imaging. (b) Momentum
distribution in one particular spin basis (other results are shown
in the supplement \cite{supplement}) for the middle flat band of
$H_{\text{CTI}}$ with open boundary condition under lattice size
$12\times12\times12$. Layers corresponding to $k_{z}=-\pi,-\pi/4,\pi/2$
are displayed. The parameters are $h=2t$, $\gamma_{T}=0.001t$, $\gamma_{P}=0.1t$. }
\label{fig:CTI-Density Distr}
\end{figure}

As an example application of our general detection method, here we
show how to measure the topological invariant $CS_{b}$ through the
TOF imaging. As shown in Fig. \ref{fig:CTI-Density Distr}(a), we
first use the TOF\ imaging to reconstruct the 3D atomic momentum
distribution. After expansion of the atomic cloud, we apply a pair
of co-propagating Raman beams focused in the $z$-axis to transfer
a layer of atoms with a fixed $z$-coordinate $z_{i}$ to another
hyperfine or Zeeman level denoted as $\left\vert r\right\rangle $.
We apply the imaging laser to couple the atoms only on the $\left\vert r\right\rangle $
level, so the imaging reads out the 2D momentum distribution $n(k_{x},k_{y},k_{z_{i}})$
with a fixed $k_{z_{i}}\varpropto z_{i}$ . We repeat this measurement
by scanning the coordinate $z_{i}$ so that each image gives a 2D
distribution $n(k_{x},k_{y},k_{z_{i}})$ with a different $k_{z_{i}}$.
By this method, we reconstruct the 3D momentum distribution $n(k_{x},k_{y},k_{z})$,
where $l$ images give $l$ pixels of $k_{z}$.

To extract the Chern-Simons term $CS_{b}$, we measure the 3D momentum
distribution $n_{bm}(k_{x},k_{y},k_{z})$ in different spin bases
to obtain the Bloch wave function $c_{bm}(\mathbf{k})$. We then use
the measured $c_{bm}(\mathbf{k})$ to calculate the gauge independent
field $\mathcal{F}_{\mu\nu}^{(b)}(\mathbf{k}_{\mathbf{J}})$ defined
in Eq. (4). By solving a discrete version of the equation $\nabla\times\mathbf{A}=\mathbf{F}$
in the momentum space with the Coulomb gauge $\nabla\cdot\mathbf{A}=0$,
we obtain the Berry connection $A_{\mu}^{(b)}(\mathbf{k}_{\mathbf{J}})$
from $\mathcal{F}_{\mu\nu}^{(b)}(\mathbf{k}_{\mathbf{J}})$. With
$A_{\mu}^{(b)}$, we extract the Chern-Simons term $CS_{b}$ using
Eq. (8).

The Chern-Simons terms extracted from our numerically simulated experiments
are shown in Table \ref{Numerics-Table} under various conditions.
Different from the Chern number case, extraction of the Chern-Simons
term using Eq. (8) does not guarantee the result to be an integer,
so the calculated values are subject to influence of numerical inaccuracies
and experimental noise. Nevertheless, from the results listed in Table
\ref{Numerics-Table}, we see that the extracted values quickly approach
the corresponding theoretical limits when we take a dozen of pixels
along each spatial dimension in the time-of-flight imaging and the
detection method remains robust to experimental imperfections (traps
and random perturbation Hamiltonians change the result by less than
$3\%$).

In summary, we have proposed a general method to experimentally measure
the topological invariants for ultracold atoms. The method is shown
to be robust to various experimental imperfections through numerically
simulated experiments.

We thank S.-L. Zhu for discussions. This work was supported by the
NBRPC (973 Program) 2011CBA00300 (2011CBA00302), the IARPA MUSIQC
program, the ARO and the AFOSR MURI program.

\onecolumngrid
\appendix
\clearpage
\section*{Supplementary Material: Direct Probe of Topological Order for Cold
Atoms}

\begin{quote}
This supplementary material gives more details on numerical simulation
of the experimental detection and extraction of the topological invariants.
In section I, we show how to numerically calculate the atomic momentum
distribution by solving the real-space Hamiltonians under open boundary
condition. In section II, we provide details on how the random perturbation
and harmonic trapping potential are incorporated into the simulation.
In Sec. III, we include more detailed results from the numerical simulations
as well as an explicit definition of the Gell-Mann matrices used in
the main text. 
\end{quote}

\section{Numerical simulation of the atomic momentum distribution}

In this section, we provide more details on how to numerically simulate
the atomic momentum distribution by solving real-space Hamiltonians.
Consider a generic quadratic Hamiltonian in real space that describe
free fermions:
\begin{eqnarray}
H & = & \sum_{\mathbf{r},\mathbf{s},\alpha,\beta}a_{\mathbf{r},\alpha}^{\dagger}\mathcal{H}_{\mathbf{r\alpha},\mathbf{s}\beta}a_{\mathbf{s},\beta},\label{eq:GeneralH}
\end{eqnarray}
where $a_{\mathbf{r},\alpha}^{\dagger}(a_{\mathbf{s},\beta})$ creates
(annihilates) a particle at lattice site $\mathbf{r}(\mathbf{s})$
with pseudospin $\alpha(\beta)$. One can solve the Schrödinger equation
$\mathcal{H}\Phi_{i}=\epsilon_{i}\Phi_{i}$ to obtain the single-particle
energy spectrum. In the matrix form, one can diagonalize $\mathcal{H}$
by a unitary transformation $U$: $\mathcal{H}=U^{\dagger}\mathcal{E}U$
to find the single-particle eigenmodes $b_{\mathbf{r},\alpha}=\sum_{\mathbf{s},\beta}U_{\mathbf{r}\alpha,\mathbf{s}\beta}a_{\mathbf{s},\beta}$.
Here $\mathcal{E}=\text{diag}(\epsilon_{1},\epsilon_{2},\cdots)$
is a diagonal matrix. For a free-fermion system described by Eq.\ (\ref{eq:GeneralH}),
the total particle number $\mathcal{N}=\sum_{\mathbf{r},\alpha}a_{\mathbf{r},\alpha}^{\dagger}a_{\mathbf{r},\alpha}$
is a conserved quantity $[H,\mathcal{N}]=0$. These $\mathcal{N}$
particles will occupy the first $\mathcal{N}$ eigenmodes with lowest
eigenenergies. Consequently, the ground state of the system reads
\begin{eqnarray}
|G\rangle & = & \prod_{i=1}^{\mathcal{N}}b_{i}^{\dagger}|0\rangle,\label{eq:GroundState}
\end{eqnarray}
where we suppress $\mathbf{r}$ and $\alpha$ into a single index
$i$ for the occupied eigenmodes, and $|0\rangle$ is the vacuum state
without any particles. The density distribution in momentum space
can then be obtained as
\begin{eqnarray}
n_{\alpha}(\mathbf{k}) & = & \langle G|a_{\alpha}^{\dagger}(\mathbf{k})a_{\alpha}(\mathbf{k})|G\rangle.\label{eq:MomentumDis}
\end{eqnarray}
where $a_{\alpha}(\mathbf{k})$ relates to $a_{\mathbf{r},\alpha}$
by a Fourier transform,
\begin{eqnarray}
n_{\alpha}(\mathbf{k}) & = & \langle G|\frac{1}{\sqrt{\mathcal{L}}}\sum_{\mathbf{r}}e^{i\mathbf{k}\cdot\mathbf{r}}a_{\mathbf{r},\alpha}^{\dagger}\frac{1}{\sqrt{\mathcal{L}}}\sum_{\mathbf{r'}}e^{-i\mathbf{k}\cdot\mathbf{r'}}a_{\mathbf{r}',\alpha}|G\rangle\nonumber \\
 & = & \frac{1}{\mathcal{L}}\sum_{\mathbf{r},\mathbf{r}'}\langle G|a_{\mathbf{r},\alpha}^{\dagger}a_{\mathbf{r}',\alpha}|G\rangle e^{i\mathbf{k}\cdot(\mathbf{r}-\mathbf{r}')},\label{eq:FourierTrans}
\end{eqnarray}
where $\mathcal{L}$ denotes the number of lattice sites. As $a_{\mathbf{r},\alpha}=\sum_{\mathbf{s},\beta}(U^{\dagger})_{\mathbf{r}\alpha,\mathbf{s}\beta}b_{\mathbf{s},\beta}$,
the quantity $\langle G|a_{\mathbf{r},\alpha}^{\dagger}a_{\mathbf{r}',\alpha}|G\rangle$
can be further simplified:
\begin{eqnarray}
\langle G|a_{\mathbf{r},\alpha}^{\dagger}a_{\mathbf{r}',\alpha}|G\rangle & = & \langle G|\sum_{\mathbf{s},\beta}(U^{\dagger})_{\mathbf{r}\alpha,\mathbf{s}\beta}^{*}b_{\mathbf{s},\beta}^{\dagger}\sum_{\mathbf{s}',\beta'}(U^{\dagger})_{\mathbf{r}'\alpha,\mathbf{s}'\beta'}b_{\mathbf{s}',\beta'}|G\rangle\label{eq:realDis}\\
 & = & \sum_{\mathbf{s},\beta,\mathbf{s}',\beta'}\langle G|b_{\mathbf{s},\beta}^{\dagger}b_{\mathbf{s}',\beta'}|G\rangle(U^{\dagger})_{\mathbf{r}\alpha,\mathbf{s}\beta}^{*}(U^{\dagger})_{\mathbf{r}'\alpha,\mathbf{s}'\beta'}\nonumber \\
 & = & \sum_{i}^{\mathcal{N}}(U^{\dagger})_{\mathbf{r}\alpha,i}^{*}(U^{\dagger})_{\mathbf{r}'\alpha,i}.\nonumber 
\end{eqnarray}
In the last step of Eq.\ (\ref{eq:realDis}), we have used the following
equation
\begin{eqnarray*}
\langle G|b_{\mathbf{s},\beta}^{\dagger}b_{\mathbf{s}',\beta'}|G\rangle & = & \begin{cases}
1 & \text{if }\mathbf{s}=\mathbf{s}',\beta=\beta'\;\text{ and the eigen-mode }b_{\mathbf{s},\beta}\text{ is occupied}\\
0 & \text{ otherwise}.
\end{cases}
\end{eqnarray*}
Combining Eq.\ (\ref{eq:realDis}) and Eq.\ (\ref{eq:FourierTrans}),
we can obtain the momentum density distribution for each pseudospin
component from a generic quadratic real-space Hamiltonian with a specific
filling fraction (the filling fraction is defined as the total particle
number divided by the lattice site number: $f=\mathcal{N}/\mathcal{L}$).
Analogously, one can rotate the pseudospin and use the same method
to compute the momentum density distribution in other spin bases $n_{a|\alpha\rangle+b|\beta\rangle}(\mathbf{k}).$

\begin{figure}
\includegraphics[width=0.6\textwidth]{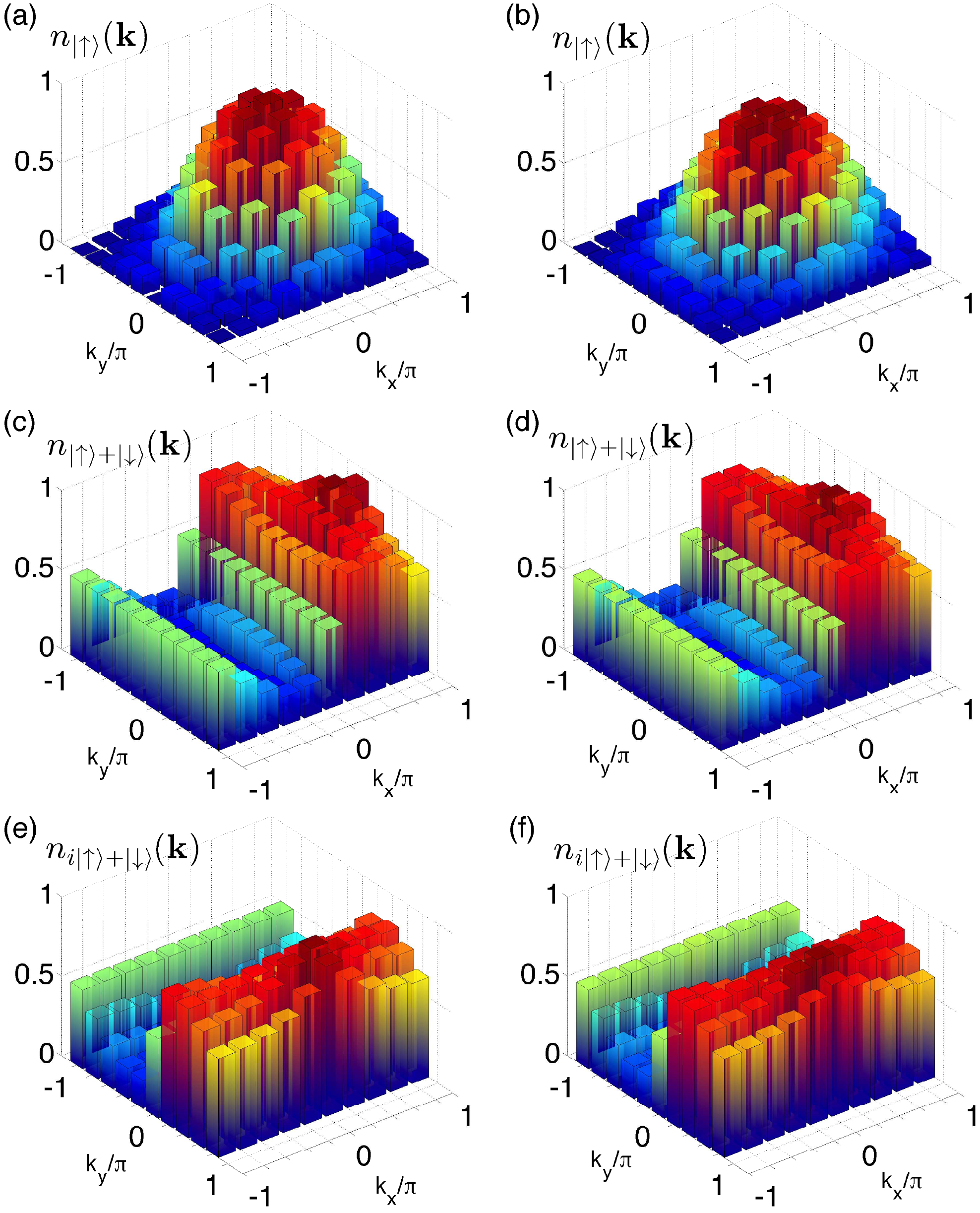}

\caption{(color online). Density distributions in momentum space for the first
band with lattice size $10\times10$. (a), (c) and (e) correspond
to the periodic boundary condition without perturbation and trapping
potential; (b), (d), and (f) correspond to open boundary conditions
with random perturbations and harmonic trapping (subfigures (b) and
(d) are repeated from the main text for clarity and completeness).
The parameters used in the calculations are chosen to be $\lambda_{\text{SO}}^{(x)}=\lambda_{\text{SO}}^{(y)}=t,$
$h=t,$ $\gamma_{\text{T}}=0.1t$, and $\gamma_{\text{P}}=0.01t$.
From these distributions, we find the Chern number $C_{1}=-1$ using
the formula in the main text. \label{suplementfig: QAH-DensityDis}}
\end{figure}

\section{Random perturbation and harmonic trapping potential}

As discussed in the main text, a typical optical lattice experiment
includes a weak harmonic trapping potential,
\begin{eqnarray}
V_{\text{T}} & = & \frac{1}{2}m_{a}\omega^{2}\sum_{\mathbf{r},\alpha}d_{\mathbf{r}}^{2}a_{\mathbf{r},\alpha}^{\dagger}a_{\mathbf{r},\alpha},\label{eq:HarmonicTrap}
\end{eqnarray}
where $d_{\mathbf{r}}$ is the distance from the center of the trap
to the lattice site $\mathbf{r}$, $m_{a}$ is the atomic mass, and
$\omega$ is the trap frequency. In our numerical simulation, we use
$\gamma_{\text{T}}=m_{a}\omega^{2}a^{2}/(2t)$ to parametrize the
influence of this trapping potential. Here $a$ is the lattice constant
and $t$ is the hopping rate. For a typical experiment, $t/\hbar\thicksim1$kHz,
$a\sim400$ nm, and $\gamma_{\text{T}}$ ranges from $10^{-3}$ ($^{6}$Li
with $\omega/2\pi=60$ Hz) to $2\times10^{-2}$ ($^{40}$K with $\omega/2\pi=100$
Hz) \cite{alba2011seeing}. To account for other possible experimental
noise, we also add a random perturbation term
\begin{eqnarray}
H_{\text{P}} & = & \gamma_{\text{P}}t\sum_{\mathbf{r},\mathbf{s},\alpha,\beta}a_{\mathbf{r},\alpha}^{\dagger}\mathcal{P}_{\mathbf{r\alpha},\mathbf{s}\beta}a_{\mathbf{s},\beta},\label{eq:RandomPert}
\end{eqnarray}
where $\mathcal{P}$ is a random Hermitian matrix with its largest
eigenvalue normalized to the unity.

In the numerical simulation, we add both $V_{\text{T}}$ and $H_{\text{P}}$
into the original Hamiltonians and calculate the momentum density
distributions using the method described in Sec.\ I. Although only
partial results are included in Fig.\ 1, Fig.\ 2 and Table I of
the main text, we have done substantial calculations with a number
of different choices of parameters $(\gamma_{\text{T}},\gamma_{\text{P}})$
for both the 2D quantum anomalous Hall effect (QAH) and 3D chiral
topological insulators (CTIs). Our results consistently show that
topological invariants extracted from the time-of-flight (TOF) measurements
are very robust to the trapping potential and random perturbations.

\begin{figure}
\includegraphics[width=0.7\textwidth]{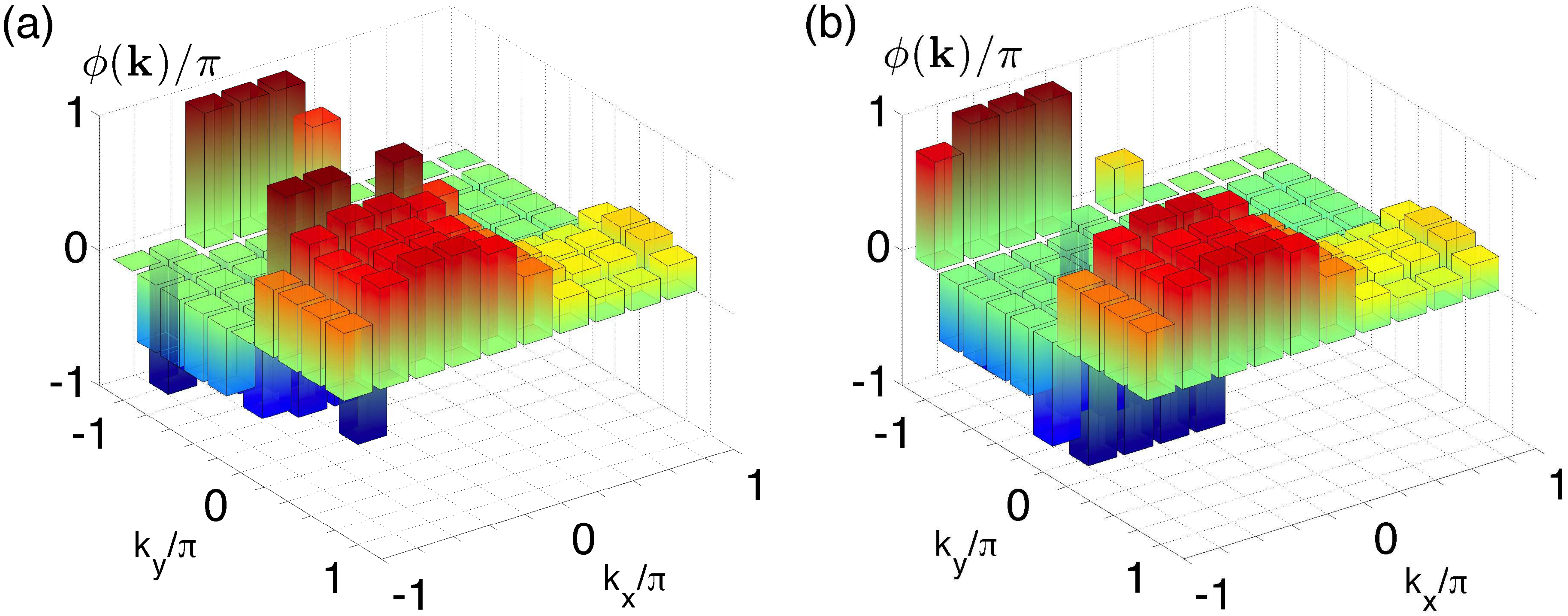}

\caption{(color online). Relative phase factors in momentum space for the first
band of the Hamiltonian $H_{\text{QAH}}$ with lattice size $10\times10$.
(a) Periodic boundary condition without perturbations and the trapping
potential. (b) Open boundary conditions with random perturbations
and a harmonic trap. The parameters are chosen to be the same as in
Fig.\ 1 of the main text. \label{suplementfig: phasefactor}}
\end{figure}

\section{More results from numerical simulation}

In this section, we provide more detailed numerical results from numerical
simulation of both the 2D QAH and 3D CTI cases.

\textit{2D QAH effect---} In Fig.\ 1 of the main text, we have plotted
two momentum density distributions of the first band. Here, we present
more plots of the density distributions in Fig.\ \ref{suplementfig: QAH-DensityDis},
considering both periodic and open boundary conditions. As discussed
in the main text, Fig.\ \ref{suplementfig: QAH-DensityDis} (b), (d)
and (f) simulate the data obtained from the TOF measurements. To extract
the Chern number, an intermediate step is to calculate the relative
phase between the spin up and down components of the Bloch wavefunction
from those density distributions. A little algebra leads to the following
equations:
\begin{eqnarray}
2n_{|\uparrow\rangle+|\downarrow\rangle}(\mathbf{k}) & = & 1+2\sqrt{n_{\uparrow}(\mathbf{k})\times(1-n_{\uparrow}(\mathbf{k}))}\cos(\phi(\mathbf{k}))\label{eq:RelativePhase1}\\
2n_{i|\uparrow\rangle+|\downarrow\rangle}(\mathbf{k}) & = & 1+2\sqrt{n_{\uparrow}(\mathbf{k})\times(1-n_{\uparrow}(\mathbf{k}))}\sin(\phi(\mathbf{k})),\label{eq:RelativePhase2}
\end{eqnarray}
where $\phi(\mathbf{k})$ is defined as the relative phase in the lower band Bloch wavefunction between the spin up and spin down components, i.e.\ $|u_{1}(\mathbf{k})\rangle= |c_{\uparrow}(\mathbf{k})|\left|\uparrow\right\rangle + |c_{\downarrow}(\mathbf{k})| e^{i\phi(\mathbf{k})}\left|\downarrow\right.\rangle$. Plugging the density distributions observed from the TOF measurements into the above equations, one obtains the relative phase. We performed the calculations for both periodic and open boundary conditions and the corresponding relative phases are shown in Fig. \ref{suplementfig: phasefactor}. With the relative phases and the density distributions, the Bloch wavefunction for the first band is determined up to a momentum-dependent overall phase. Using the method introduced in the main text, we are able to extract the desired Chern number $C_{1}=-1$.

\begin{figure}
\includegraphics[width=0.9\textwidth]{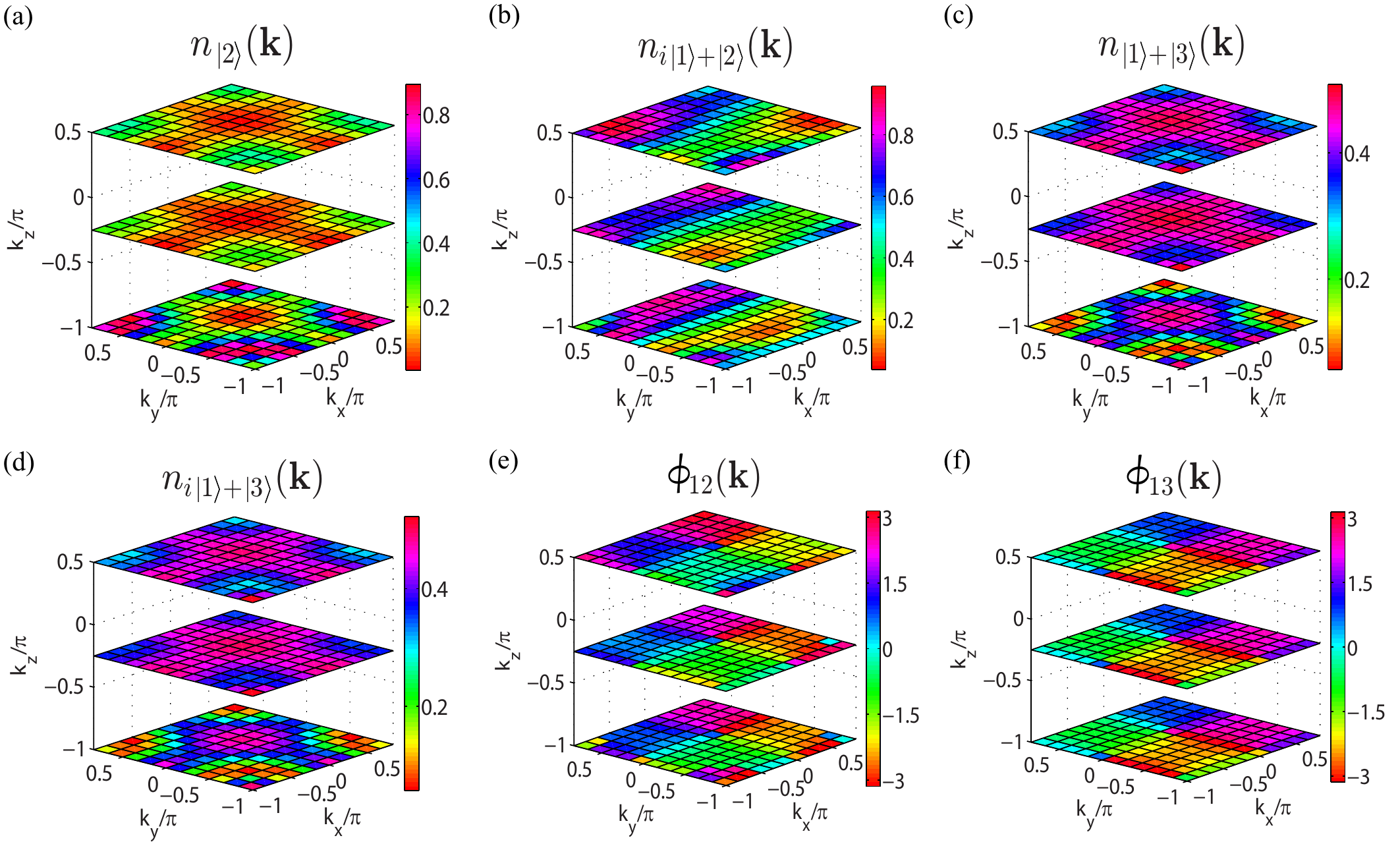}

\caption{(color online). Momentum density distributions and relative phase
factors for the middle flat band with open boundary conditions for
$H_{\text{CTI}}$ including a harmonic trap and some random perturbations.
The lattice size is $12\times12\times12$. Layers corresponding to
$k_{z}=-\pi,-\pi/4,\pi/2$ are displayed. $\phi_{12}(\mathbf{k})$ ($\phi_{13}(\mathbf{k})$)  is the phase factor between spin 1 and 2 (spin 1 and 3) in the Bloch wavefunction. The parameters are chosen to be the same as in Fig.\ 2 of the main text.\label{supplementfig: CTI}}
\end{figure}

\textit{3D CTI---} Let us first write down explicitly the four Gell-Mann
matrices used in the Hamiltonian $H_{\text{CTI}}$ in the main text:

\begin{eqnarray*}
G_{4} & = & \left(\begin{array}{ccc}
0 & 0 & 1\\
0 & 0 & 0\\
1 & 0 & 0
\end{array}\right),\quad G_{5}=\left(\begin{array}{ccc}
0 & 0 & -i\\
0 & 0 & 0\\
i & 0 & 0
\end{array}\right),\quad G_{6}=\left(\begin{array}{ccc}
0 & 0 & 0\\
0 & 0 & 1\\
0 & 1 & 0
\end{array}\right),\quad G_{7}=\left(\begin{array}{ccc}
0 & 0 & 0\\
0 & 0 & -i\\
0 & i & 0
\end{array}\right).
\end{eqnarray*}
A Fourier transform brings $H_{\text{CTI}}$ to the momentum space
\cite{Wang2014probe, neupert2012noncommutative}:
\begin{eqnarray*}
H_{\text{CTI}} & = & \sum_{\mathbf{k}}\psi_{\mathbf{k}}^{\dagger}\mathcal{H}_{\text{CTI}}(\mathbf{k})\psi_{\mathbf{k}},
\end{eqnarray*}
where $\psi_{\mathbf{k}}^{\dagger}=(a_{\mathbf{k},1}^{\dagger},a_{\mathbf{k},2}^{\dagger},a_{\mathbf{k},3}^{\dagger})$
and $\mathcal{H}_{\text{CTI}}(\mathbf{k})=\sum_{j=1}^{4}G_{3+j}q_{j}(\mathbf{k})$
with $[q_{1}(\mathbf{k}),q_{2}(\mathbf{k}),q_{3}(\mathbf{k}),q_{4}(\mathbf{k})]=[\sin k_{x},\sin k_{y},\sin k_{z},m-\cos k_{x}-\cos k_{y}-\cos k_{z}]$.
One can easily check that this Hamiltonian indeed has a chiral symmetry
represented by $S\mathcal{H}_{\text{CTI}}(\mathbf{k})S^{-1}=-\mathcal{H}_{\text{CTI}}(\mathbf{k})$,
where $S\equiv\text{diag}(1,1,-1)$ is a unitary matrix. This chiral
symmetry leads to an exact zero-energy flat band as discussed in Ref.\ \cite{neupert2012noncommutative,Wang2014probe}.

In the main text, we plotted some of the momentum density distributions
of the middle flat band in Fig.\ 2(b). Here we include more results
of the density distributions and the relative phases in Fig. \ref{supplementfig: CTI}.
Analogous to the case of QAH effect, the momentum density distributions
can be directly observed from the layered TOF measurements and the
relative phases can be calculated from the observed density distributions.
After all density distributions are observed in an actual experiment,
the Chern-Simons term characterizing the topological structure of
the Bloch band can be readily extracted with the method described
in the main text.

Besides the Chern-Simons term for the middle flat band shown in Table
I of the main text, we have also calculated it for both the first
(lowest) and third (highest) bands. Our results are recorded in Table
I here. From this table, the extracted Chern-Simons terms converge
to the expected theoretical value $CS_{1}/\pi=CS_{3}/\pi=1/4$ as
we increase the lattice size.

\begin{table}
\caption{The Chern-Simons terms of the first and third bands for the Hamiltonian
$H_{\text{CTI}}$. The parameters are chosen to be the same as in
Fig.\ 2b of the main text.}

\begin{ruledtabular}%
\begin{tabular}{ccccccc}
 & Size  & $h/t$  & Periodic  & Open  & Trap  & Pert.+Trap \tabularnewline
\hline
CTI ($CS_{1}/\pi$)  & $10^{3}$  & 2  & 0.246  & 0.228  & 0.231  & 0.231 \tabularnewline
 & $12^{3}$  & 2  & 0.248  & 0.228  & 0.235  & 0.235\tabularnewline
 & $10^{3}$  & 4  & $5.8\times10^{-5}$  & $8.6\times10^{-5}$  & $1.5\times10^{-4}$  & $1.4\times10^{-4}$\tabularnewline
 &  &  &  &  &  & \tabularnewline
CTI ($CS_{3}/\pi$)  & $10^{3}$  & 2  & 0.246  & $0.226$  & $0.227$  & $0.227$ \tabularnewline
 & $12^{3}$  & 2  & $0.248$  & $0.229$  & $0.230$  & $0.231$ \tabularnewline
 & $10^{3}$  & 4  & $5.8\times10^{-5}$  & $5.0\times10^{-5}$  & $1.7\times10^{-4}$  & $1.8\times10^{-4}$\tabularnewline
\end{tabular}\end{ruledtabular} \label{Numerics-Table}
\end{table}


\begin{thebibliography}{61}%
\makeatletter
\providecommand \@ifxundefined [1]{%
 \@ifx{#1\undefined}
}%
\providecommand \@ifnum [1]{%
 \ifnum #1\expandafter \@firstoftwo
 \else \expandafter \@secondoftwo
 \fi
}%
\providecommand \@ifx [1]{%
 \ifx #1\expandafter \@firstoftwo
 \else \expandafter \@secondoftwo
 \fi
}%
\providecommand \natexlab [1]{#1}%
\providecommand \enquote  [1]{``#1''}%
\providecommand \bibnamefont  [1]{#1}%
\providecommand \bibfnamefont [1]{#1}%
\providecommand \citenamefont [1]{#1}%
\providecommand \href@noop [0]{\@secondoftwo}%
\providecommand \href [0]{\begingroup \@sanitize@url \@href}%
\providecommand \@href[1]{\@@startlink{#1}\@@href}%
\providecommand \@@href[1]{\endgroup#1\@@endlink}%
\providecommand \@sanitize@url [0]{\catcode `\\12\catcode `\$12\catcode
  `\&12\catcode `\#12\catcode `\^12\catcode `\_12\catcode `\%12\relax}%
\providecommand \@@startlink[1]{}%
\providecommand \@@endlink[0]{}%
\providecommand \url  [0]{\begingroup\@sanitize@url \@url }%
\providecommand \@url [1]{\endgroup\@href {#1}{\urlprefix }}%
\providecommand \urlprefix  [0]{URL }%
\providecommand \Eprint [0]{\href }%
\providecommand \doibase [0]{http://dx.doi.org/}%
\providecommand \selectlanguage [0]{\@gobble}%
\providecommand \bibinfo  [0]{\@secondoftwo}%
\providecommand \bibfield  [0]{\@secondoftwo}%
\providecommand \translation [1]{[#1]}%
\providecommand \BibitemOpen [0]{}%
\providecommand \bibitemStop [0]{}%
\providecommand \bibitemNoStop [0]{.\EOS\space}%
\providecommand \EOS [0]{\spacefactor3000\relax}%
\providecommand \BibitemShut  [1]{\csname bibitem#1\endcsname}%
\let\auto@bib@innerbib\@empty
\bibitem [{\citenamefont {Hasan}\ and\ \citenamefont
  {Kane}(2010)}]{hasan2010colloquium}%
  \BibitemOpen
  \bibfield  {author} {\bibinfo {author} {\bibfnamefont {M.~Z.}\ \bibnamefont
  {Hasan}}\ and\ \bibinfo {author} {\bibfnamefont {C.~L.}\ \bibnamefont
  {Kane}},\ }\href@noop {} {\bibfield  {journal} {\bibinfo  {journal} {Rev.
  Mod. Phys.}\ }\textbf {\bibinfo {volume} {82}},\ \bibinfo {pages} {3045}
  (\bibinfo {year} {2010})}\BibitemShut {NoStop}%
\bibitem [{\citenamefont {Qi}\ and\ \citenamefont
  {Zhang}(2011)}]{qi2011topological}%
  \BibitemOpen
  \bibfield  {author} {\bibinfo {author} {\bibfnamefont {X.-L.}\ \bibnamefont
  {Qi}}\ and\ \bibinfo {author} {\bibfnamefont {S.-C.}\ \bibnamefont {Zhang}},\
  }\href@noop {} {\bibfield  {journal} {\bibinfo  {journal} {Rev. Mod. Phys.}\
  }\textbf {\bibinfo {volume} {83}},\ \bibinfo {pages} {1057} (\bibinfo {year}
  {2011})}\BibitemShut {NoStop}%
\bibitem [{\citenamefont {Moore}(2010)}]{moore2010birth}%
  \BibitemOpen
  \bibfield  {author} {\bibinfo {author} {\bibfnamefont {J.~E.}\ \bibnamefont
  {Moore}},\ }\href@noop {} {\bibfield  {journal} {\bibinfo  {journal}
  {Nature}\ }\textbf {\bibinfo {volume} {464}},\ \bibinfo {pages} {194}
  (\bibinfo {year} {2010})}\BibitemShut {NoStop}%
\bibitem [{\citenamefont {Schnyder}\ \emph {et~al.}(2008)\citenamefont
  {Schnyder}, \citenamefont {Ryu}, \citenamefont {Furusaki},\ and\
  \citenamefont {Ludwig}}]{schnyder2008classification}%
  \BibitemOpen
  \bibfield  {author} {\bibinfo {author} {\bibfnamefont {A.~P.}\ \bibnamefont
  {Schnyder}}, \bibinfo {author} {\bibfnamefont {S.}~\bibnamefont {Ryu}},
  \bibinfo {author} {\bibfnamefont {A.}~\bibnamefont {Furusaki}}, \ and\
  \bibinfo {author} {\bibfnamefont {A.~W.}\ \bibnamefont {Ludwig}},\
  }\href@noop {} {\bibfield  {journal} {\bibinfo  {journal} {Phys. Rev. B}\
  }\textbf {\bibinfo {volume} {78}},\ \bibinfo {pages} {195125} (\bibinfo
  {year} {2008})}\BibitemShut {NoStop}%
\bibitem [{\citenamefont {Kitaev}(2009)}]{kitaev2009periodic}%
  \BibitemOpen
  \bibfield  {author} {\bibinfo {author} {\bibfnamefont {A.}~\bibnamefont
  {Kitaev}},\ }in\ \href@noop {} {\emph {\bibinfo {booktitle} {AIP Conference
  Proceedings}}},\ Vol.\ \bibinfo {volume} {1134}\ (\bibinfo {year} {2009})\
  p.~\bibinfo {pages} {22}\BibitemShut {NoStop}%
\bibitem [{\citenamefont {Ryu}\ \emph {et~al.}(2010)\citenamefont {Ryu},
  \citenamefont {Schnyder}, \citenamefont {Furusaki},\ and\ \citenamefont
  {Ludwig}}]{ryu2010topological}%
  \BibitemOpen
  \bibfield  {author} {\bibinfo {author} {\bibfnamefont {S.}~\bibnamefont
  {Ryu}}, \bibinfo {author} {\bibfnamefont {A.~P.}\ \bibnamefont {Schnyder}},
  \bibinfo {author} {\bibfnamefont {A.}~\bibnamefont {Furusaki}}, \ and\
  \bibinfo {author} {\bibfnamefont {A.~W.}\ \bibnamefont {Ludwig}},\
  }\href@noop {} {\bibfield  {journal} {\bibinfo  {journal} {New J. Phys.}\
  }\textbf {\bibinfo {volume} {12}},\ \bibinfo {pages} {065010} (\bibinfo
  {year} {2010})}\BibitemShut {NoStop}%
\bibitem [{\citenamefont {Thouless}\ \emph {et~al.}(1982)\citenamefont
  {Thouless}, \citenamefont {Kohmoto}, \citenamefont {Nightingale},\ and\
  \citenamefont {Den~Nijs}}]{thouless1982quantized}%
  \BibitemOpen
  \bibfield  {author} {\bibinfo {author} {\bibfnamefont {D.}~\bibnamefont
  {Thouless}}, \bibinfo {author} {\bibfnamefont {M.}~\bibnamefont {Kohmoto}},
  \bibinfo {author} {\bibfnamefont {M.}~\bibnamefont {Nightingale}}, \ and\
  \bibinfo {author} {\bibfnamefont {M.}~\bibnamefont {Den~Nijs}},\ }\href@noop
  {} {\bibfield  {journal} {\bibinfo  {journal} {Phys. Rev. Lett.}\ }\textbf
  {\bibinfo {volume} {49}},\ \bibinfo {pages} {405} (\bibinfo {year}
  {1982})}\BibitemShut {NoStop}%
\bibitem [{\citenamefont {von Klitzing}(1986)}]{von1986quantized}%
  \BibitemOpen
  \bibfield  {author} {\bibinfo {author} {\bibfnamefont {K.}~\bibnamefont {von
  Klitzing}},\ }\href@noop {} {\bibfield  {journal} {\bibinfo  {journal} {Rev.
  Mod. Phys.}\ }\textbf {\bibinfo {volume} {58}} (\bibinfo {year}
  {1986})}\BibitemShut {NoStop}%
\bibitem [{\citenamefont {Sarma}\ and\ \citenamefont
  {Pinczuk}(2008)}]{sarma2008perspectives}%
  \BibitemOpen
  \bibfield  {author} {\bibinfo {author} {\bibfnamefont {S.~D.}\ \bibnamefont
  {Sarma}}\ and\ \bibinfo {author} {\bibfnamefont {A.}~\bibnamefont
  {Pinczuk}},\ }\href@noop {} {\emph {\bibinfo {title} {Perspectives in quantum
  Hall effects}}}\ (\bibinfo  {publisher} {John Wiley \& Sons},\ \bibinfo
  {year} {2008})\BibitemShut {NoStop}%
\bibitem [{\citenamefont {Lin}\ \emph {et~al.}(2009)\citenamefont {Lin},
  \citenamefont {Compton}, \citenamefont {Jimenez-Garcia}, \citenamefont
  {Porto},\ and\ \citenamefont {Spielman}}]{lin2009synthetic}%
  \BibitemOpen
  \bibfield  {author} {\bibinfo {author} {\bibfnamefont {Y.-J.}\ \bibnamefont
  {Lin}}, \bibinfo {author} {\bibfnamefont {R.~L.}\ \bibnamefont {Compton}},
  \bibinfo {author} {\bibfnamefont {K.}~\bibnamefont {Jimenez-Garcia}},
  \bibinfo {author} {\bibfnamefont {J.~V.}\ \bibnamefont {Porto}}, \ and\
  \bibinfo {author} {\bibfnamefont {I.~B.}\ \bibnamefont {Spielman}},\
  }\href@noop {} {\bibfield  {journal} {\bibinfo  {journal} {Nature}\ }\textbf
  {\bibinfo {volume} {462}},\ \bibinfo {pages} {628} (\bibinfo {year}
  {2009})}\BibitemShut {NoStop}%
\bibitem [{\citenamefont {Liu}\ \emph {et~al.}(2009)\citenamefont {Liu},
  \citenamefont {Borunda}, \citenamefont {Liu},\ and\ \citenamefont
  {Sinova}}]{liu2009effect}%
  \BibitemOpen
  \bibfield  {author} {\bibinfo {author} {\bibfnamefont {X.-J.}\ \bibnamefont
  {Liu}}, \bibinfo {author} {\bibfnamefont {M.~F.}\ \bibnamefont {Borunda}},
  \bibinfo {author} {\bibfnamefont {X.}~\bibnamefont {Liu}}, \ and\ \bibinfo
  {author} {\bibfnamefont {J.}~\bibnamefont {Sinova}},\ }\href@noop {}
  {\bibfield  {journal} {\bibinfo  {journal} {Phys. Rev. Lett.}\ }\textbf
  {\bibinfo {volume} {102}},\ \bibinfo {pages} {046402} (\bibinfo {year}
  {2009})}\BibitemShut {NoStop}%
\bibitem [{\citenamefont {Wang}\ \emph {et~al.}(2012)\citenamefont {Wang},
  \citenamefont {Yu}, \citenamefont {Fu}, \citenamefont {Miao}, \citenamefont
  {Huang}, \citenamefont {Chai}, \citenamefont {Zhai},\ and\ \citenamefont
  {Zhang}}]{wang2012spin}%
  \BibitemOpen
  \bibfield  {author} {\bibinfo {author} {\bibfnamefont {P.}~\bibnamefont
  {Wang}}, \bibinfo {author} {\bibfnamefont {Z.-Q.}\ \bibnamefont {Yu}},
  \bibinfo {author} {\bibfnamefont {Z.}~\bibnamefont {Fu}}, \bibinfo {author}
  {\bibfnamefont {J.}~\bibnamefont {Miao}}, \bibinfo {author} {\bibfnamefont
  {L.}~\bibnamefont {Huang}}, \bibinfo {author} {\bibfnamefont
  {S.}~\bibnamefont {Chai}}, \bibinfo {author} {\bibfnamefont {H.}~\bibnamefont
  {Zhai}}, \ and\ \bibinfo {author} {\bibfnamefont {J.}~\bibnamefont {Zhang}},\
  }\href@noop {} {\bibfield  {journal} {\bibinfo  {journal} {Phys. Rev. Lett.}\
  }\textbf {\bibinfo {volume} {109}},\ \bibinfo {pages} {095301} (\bibinfo
  {year} {2012})}\BibitemShut {NoStop}%
\bibitem [{\citenamefont {Cheuk}\ \emph {et~al.}(2012)\citenamefont {Cheuk},
  \citenamefont {Sommer}, \citenamefont {Hadzibabic}, \citenamefont {Yefsah},
  \citenamefont {Bakr},\ and\ \citenamefont {Zwierlein}}]{cheuk2012spin}%
  \BibitemOpen
  \bibfield  {author} {\bibinfo {author} {\bibfnamefont {L.~W.}\ \bibnamefont
  {Cheuk}}, \bibinfo {author} {\bibfnamefont {A.~T.}\ \bibnamefont {Sommer}},
  \bibinfo {author} {\bibfnamefont {Z.}~\bibnamefont {Hadzibabic}}, \bibinfo
  {author} {\bibfnamefont {T.}~\bibnamefont {Yefsah}}, \bibinfo {author}
  {\bibfnamefont {W.~S.}\ \bibnamefont {Bakr}}, \ and\ \bibinfo {author}
  {\bibfnamefont {M.~W.}\ \bibnamefont {Zwierlein}},\ }\href@noop {} {\bibfield
   {journal} {\bibinfo  {journal} {Phys. Rev. Lett.}\ }\textbf {\bibinfo
  {volume} {109}},\ \bibinfo {pages} {095302} (\bibinfo {year}
  {2012})}\BibitemShut {NoStop}%
\bibitem [{\citenamefont {Dalibard}\ \emph {et~al.}(2011)\citenamefont
  {Dalibard}, \citenamefont {Gerbier}, \citenamefont {Juzeli{\=u}nas},\ and\
  \citenamefont {{\"O}hberg}}]{dalibard2011colloquium}%
  \BibitemOpen
  \bibfield  {author} {\bibinfo {author} {\bibfnamefont {J.}~\bibnamefont
  {Dalibard}}, \bibinfo {author} {\bibfnamefont {F.}~\bibnamefont {Gerbier}},
  \bibinfo {author} {\bibfnamefont {G.}~\bibnamefont {Juzeli{\=u}nas}}, \ and\
  \bibinfo {author} {\bibfnamefont {P.}~\bibnamefont {{\"O}hberg}},\
  }\href@noop {} {\bibfield  {journal} {\bibinfo  {journal} {Rev. Mod. Phys.}\
  }\textbf {\bibinfo {volume} {83}},\ \bibinfo {pages} {1523} (\bibinfo {year}
  {2011})}\BibitemShut {NoStop}%
\bibitem [{\citenamefont {Galitski}\ and\ \citenamefont
  {Spielman}(2013)}]{galitski2013spin}%
  \BibitemOpen
  \bibfield  {author} {\bibinfo {author} {\bibfnamefont {V.}~\bibnamefont
  {Galitski}}\ and\ \bibinfo {author} {\bibfnamefont {I.~B.}\ \bibnamefont
  {Spielman}},\ }\href@noop {} {\bibfield  {journal} {\bibinfo  {journal}
  {Nature}\ }\textbf {\bibinfo {volume} {494}},\ \bibinfo {pages} {49}
  (\bibinfo {year} {2013})}\BibitemShut {NoStop}%
\bibitem [{\citenamefont {Lewenstein}\ \emph {et~al.}(2007)\citenamefont
  {Lewenstein}, \citenamefont {Sanpera}, \citenamefont {Ahufinger},
  \citenamefont {Damski}, \citenamefont {Sen},\ and\ \citenamefont
  {Sen}}]{lewenstein2007ultracold}%
  \BibitemOpen
  \bibfield  {author} {\bibinfo {author} {\bibfnamefont {M.}~\bibnamefont
  {Lewenstein}}, \bibinfo {author} {\bibfnamefont {A.}~\bibnamefont {Sanpera}},
  \bibinfo {author} {\bibfnamefont {V.}~\bibnamefont {Ahufinger}}, \bibinfo
  {author} {\bibfnamefont {B.}~\bibnamefont {Damski}}, \bibinfo {author}
  {\bibfnamefont {A.}~\bibnamefont {Sen}}, \ and\ \bibinfo {author}
  {\bibfnamefont {U.}~\bibnamefont {Sen}},\ }\href@noop {} {\bibfield
  {journal} {\bibinfo  {journal} {Adv. Phys.}\ }\textbf {\bibinfo {volume}
  {56}},\ \bibinfo {pages} {243} (\bibinfo {year} {2007})}\BibitemShut
  {NoStop}%
\bibitem [{\citenamefont {Bloch}\ \emph {et~al.}(2012)\citenamefont {Bloch},
  \citenamefont {Dalibard},\ and\ \citenamefont
  {Nascimb{\`e}ne}}]{bloch2012quantum}%
  \BibitemOpen
  \bibfield  {author} {\bibinfo {author} {\bibfnamefont {I.}~\bibnamefont
  {Bloch}}, \bibinfo {author} {\bibfnamefont {J.}~\bibnamefont {Dalibard}}, \
  and\ \bibinfo {author} {\bibfnamefont {S.}~\bibnamefont {Nascimb{\`e}ne}},\
  }\href@noop {} {\bibfield  {journal} {\bibinfo  {journal} {Nat. Phys.}\
  }\textbf {\bibinfo {volume} {8}},\ \bibinfo {pages} {267} (\bibinfo {year}
  {2012})}\BibitemShut {NoStop}%
\bibitem [{\citenamefont {Zhu}\ \emph {et~al.}(2006)\citenamefont {Zhu},
  \citenamefont {Fu}, \citenamefont {Wu}, \citenamefont {Zhang},\ and\
  \citenamefont {Duan}}]{zhu2006spin}%
  \BibitemOpen
  \bibfield  {author} {\bibinfo {author} {\bibfnamefont {S.-L.}\ \bibnamefont
  {Zhu}}, \bibinfo {author} {\bibfnamefont {H.}~\bibnamefont {Fu}}, \bibinfo
  {author} {\bibfnamefont {C.-J.}\ \bibnamefont {Wu}}, \bibinfo {author}
  {\bibfnamefont {S.-C.}\ \bibnamefont {Zhang}}, \ and\ \bibinfo {author}
  {\bibfnamefont {L.-M.}\ \bibnamefont {Duan}},\ }\href@noop {} {\bibfield
  {journal} {\bibinfo  {journal} {Phys. Rev. Lett.}\ }\textbf {\bibinfo
  {volume} {97}},\ \bibinfo {pages} {240401} (\bibinfo {year}
  {2006})}\BibitemShut {NoStop}%
\bibitem [{\citenamefont {Zhu}\ \emph {et~al.}(2011)\citenamefont {Zhu},
  \citenamefont {Shao}, \citenamefont {Wang},\ and\ \citenamefont
  {Duan}}]{zhu2011probing}%
  \BibitemOpen
  \bibfield  {author} {\bibinfo {author} {\bibfnamefont {S.-L.}\ \bibnamefont
  {Zhu}}, \bibinfo {author} {\bibfnamefont {L.-B.}\ \bibnamefont {Shao}},
  \bibinfo {author} {\bibfnamefont {Z.}~\bibnamefont {Wang}}, \ and\ \bibinfo
  {author} {\bibfnamefont {L.-M.}\ \bibnamefont {Duan}},\ }\href@noop {}
  {\bibfield  {journal} {\bibinfo  {journal} {Phys. Rev. Lett.}\ }\textbf
  {\bibinfo {volume} {106}},\ \bibinfo {pages} {100404} (\bibinfo {year}
  {2011})}\BibitemShut {NoStop}%
\bibitem [{\citenamefont {B{\'e}ri}\ and\ \citenamefont
  {Cooper}(2011)}]{beri2011z_}%
  \BibitemOpen
  \bibfield  {author} {\bibinfo {author} {\bibfnamefont {B.}~\bibnamefont
  {B{\'e}ri}}\ and\ \bibinfo {author} {\bibfnamefont {N.}~\bibnamefont
  {Cooper}},\ }\href@noop {} {\bibfield  {journal} {\bibinfo  {journal} {Phys.
  Rev. Lett.}\ }\textbf {\bibinfo {volume} {107}},\ \bibinfo {pages} {145301}
  (\bibinfo {year} {2011})}\BibitemShut {NoStop}%
\bibitem [{\citenamefont {Aidelsburger}\ \emph {et~al.}(2013)\citenamefont
  {Aidelsburger}, \citenamefont {Atala}, \citenamefont {Lohse}, \citenamefont
  {Barreiro}, \citenamefont {Paredes},\ and\ \citenamefont
  {Bloch}}]{aidelsburger2013realization}%
  \BibitemOpen
  \bibfield  {author} {\bibinfo {author} {\bibfnamefont {M.}~\bibnamefont
  {Aidelsburger}}, \bibinfo {author} {\bibfnamefont {M.}~\bibnamefont {Atala}},
  \bibinfo {author} {\bibfnamefont {M.}~\bibnamefont {Lohse}}, \bibinfo
  {author} {\bibfnamefont {J.}~\bibnamefont {Barreiro}}, \bibinfo {author}
  {\bibfnamefont {B.}~\bibnamefont {Paredes}}, \ and\ \bibinfo {author}
  {\bibfnamefont {I.}~\bibnamefont {Bloch}},\ }\href@noop {} {\bibfield
  {journal} {\bibinfo  {journal} {Phys. Rev. Lett.}\ }\textbf {\bibinfo
  {volume} {111}},\ \bibinfo {pages} {185301} (\bibinfo {year}
  {2013})}\BibitemShut {NoStop}%
\bibitem [{\citenamefont {Miyake}\ \emph {et~al.}(2013)\citenamefont {Miyake},
  \citenamefont {Siviloglou}, \citenamefont {Kennedy}, \citenamefont {Burton},\
  and\ \citenamefont {Ketterle}}]{miyake2013realizing}%
  \BibitemOpen
  \bibfield  {author} {\bibinfo {author} {\bibfnamefont {H.}~\bibnamefont
  {Miyake}}, \bibinfo {author} {\bibfnamefont {G.~A.}\ \bibnamefont
  {Siviloglou}}, \bibinfo {author} {\bibfnamefont {C.~J.}\ \bibnamefont
  {Kennedy}}, \bibinfo {author} {\bibfnamefont {W.~C.}\ \bibnamefont {Burton}},
  \ and\ \bibinfo {author} {\bibfnamefont {W.}~\bibnamefont {Ketterle}},\
  }\href@noop {} {\bibfield  {journal} {\bibinfo  {journal} {Phys. Rev. Lett.}\
  }\textbf {\bibinfo {volume} {111}},\ \bibinfo {pages} {185302} (\bibinfo
  {year} {2013})}\BibitemShut {NoStop}%
\bibitem [{\citenamefont {Shao}\ \emph {et~al.}(2008)\citenamefont {Shao},
  \citenamefont {Zhu}, \citenamefont {Sheng}, \citenamefont {Xing},\ and\
  \citenamefont {Wang}}]{shao2008realizing}%
  \BibitemOpen
  \bibfield  {author} {\bibinfo {author} {\bibfnamefont {L.}~\bibnamefont
  {Shao}}, \bibinfo {author} {\bibfnamefont {S.-L.}\ \bibnamefont {Zhu}},
  \bibinfo {author} {\bibfnamefont {L.}~\bibnamefont {Sheng}}, \bibinfo
  {author} {\bibfnamefont {D.}~\bibnamefont {Xing}}, \ and\ \bibinfo {author}
  {\bibfnamefont {Z.}~\bibnamefont {Wang}},\ }\href@noop {} {\bibfield
  {journal} {\bibinfo  {journal} {Phys. Rev. Lett.}\ }\textbf {\bibinfo
  {volume} {101}},\ \bibinfo {pages} {246810} (\bibinfo {year}
  {2008})}\BibitemShut {NoStop}%
\bibitem [{\citenamefont {Dauphin}\ and\ \citenamefont
  {Goldman}(2013)}]{dauphin2013extracting}%
  \BibitemOpen
  \bibfield  {author} {\bibinfo {author} {\bibfnamefont {A.}~\bibnamefont
  {Dauphin}}\ and\ \bibinfo {author} {\bibfnamefont {N.}~\bibnamefont
  {Goldman}},\ }\href@noop {} {\bibfield  {journal} {\bibinfo  {journal} {Phys.
  Rev. Lett.}\ }\textbf {\bibinfo {volume} {111}},\ \bibinfo {pages} {135302}
  (\bibinfo {year} {2013})}\BibitemShut {NoStop}%
   \bibitem{Wang2013ChargeP}
 L. Wang, A. A. Soluyanov, and M. Troyer, Phys. Rev. Lett. \textbf{110}, 166802 (2013).
  \bibitem [{\citenamefont {Hauke}\ \emph {et~al.}(2014)\citenamefont {Hauke},
  \citenamefont {Lewenstein},\ and\ \citenamefont
  {Eckardt}}]{Hauke2014Tomography}%
  \BibitemOpen
  \bibfield  {author} {\bibinfo {author} {\bibfnamefont {P.}~\bibnamefont
  {Hauke}}, \bibinfo {author} {\bibfnamefont {M.}~\bibnamefont {Lewenstein}}, \
  and\ \bibinfo {author} {\bibfnamefont {A.}~\bibnamefont {Eckardt}},\ }\href
  {\doibase 10.1103/PhysRevLett.113.045303} {\bibfield  {journal} {\bibinfo
  {journal} {Phys. Rev. Lett.}\ }\textbf {\bibinfo {volume} {113}},\ \bibinfo
  {pages} {045303} (\bibinfo {year} {2014})}\BibitemShut {NoStop}%
\bibitem [{\citenamefont {Liu}\ \emph {et~al.}(2010)\citenamefont {Liu},
  \citenamefont {Liu}, \citenamefont {Wu},\ and\ \citenamefont
  {Sinova}}]{liu2010quantum}%
  \BibitemOpen
  \bibfield  {author} {\bibinfo {author} {\bibfnamefont {X.-J.}\ \bibnamefont
  {Liu}}, \bibinfo {author} {\bibfnamefont {X.}~\bibnamefont {Liu}}, \bibinfo
  {author} {\bibfnamefont {C.}~\bibnamefont {Wu}}, \ and\ \bibinfo {author}
  {\bibfnamefont {J.}~\bibnamefont {Sinova}},\ }\href@noop {} {\bibfield
  {journal} {\bibinfo  {journal} {Phys. Rev. A}\ }\textbf {\bibinfo {volume}
  {81}},\ \bibinfo {pages} {033622} (\bibinfo {year} {2010})}\BibitemShut
  {NoStop}%
\bibitem [{\citenamefont {Goldman}\ \emph {et~al.}(2012)\citenamefont
  {Goldman}, \citenamefont {Beugnon},\ and\ \citenamefont
  {Gerbier}}]{goldman2012detecting}%
  \BibitemOpen
  \bibfield  {author} {\bibinfo {author} {\bibfnamefont {N.}~\bibnamefont
  {Goldman}}, \bibinfo {author} {\bibfnamefont {J.}~\bibnamefont {Beugnon}}, \
  and\ \bibinfo {author} {\bibfnamefont {F.}~\bibnamefont {Gerbier}},\
  }\href@noop {} {\bibfield  {journal} {\bibinfo  {journal} {Phys. Rev. Lett.}\
  }\textbf {\bibinfo {volume} {108}},\ \bibinfo {pages} {255303} (\bibinfo
  {year} {2012})}\BibitemShut {NoStop}%
\bibitem [{\citenamefont {Goldman}\ \emph {et~al.}(2013)\citenamefont
  {Goldman}, \citenamefont {Dalibard}, \citenamefont {Dauphin}, \citenamefont
  {Gerbier}, \citenamefont {Lewenstein}, \citenamefont {Zoller},\ and\
  \citenamefont {Spielman}}]{goldman2013direct}%
  \BibitemOpen
  \bibfield  {author} {\bibinfo {author} {\bibfnamefont {N.}~\bibnamefont
  {Goldman}}, \bibinfo {author} {\bibfnamefont {J.}~\bibnamefont {Dalibard}},
  \bibinfo {author} {\bibfnamefont {A.}~\bibnamefont {Dauphin}}, \bibinfo
  {author} {\bibfnamefont {F.}~\bibnamefont {Gerbier}}, \bibinfo {author}
  {\bibfnamefont {M.}~\bibnamefont {Lewenstein}}, \bibinfo {author}
  {\bibfnamefont {P.}~\bibnamefont {Zoller}}, \ and\ \bibinfo {author}
  {\bibfnamefont {I.~B.}\ \bibnamefont {Spielman}},\ }\href@noop {} {\bibfield
  {journal} {\bibinfo  {journal} {Proc. Natl. Acad. Sci.}\ }\textbf {\bibinfo
  {volume} {110}},\ \bibinfo {pages} {6736} (\bibinfo {year}
  {2013})}\BibitemShut {NoStop}%
\bibitem [{\citenamefont {Zhao}\ \emph {et~al.}(2011)\citenamefont {Zhao},
  \citenamefont {Bray-Ali}, \citenamefont {Williams}, \citenamefont
  {Spielman},\ and\ \citenamefont {Satija}}]{Zhao2011Chern}%
  \BibitemOpen
  \bibfield  {author} {\bibinfo {author} {\bibfnamefont {E.}~\bibnamefont
  {Zhao}}, \bibinfo {author} {\bibfnamefont {N.}~\bibnamefont {Bray-Ali}},
  \bibinfo {author} {\bibfnamefont {C.~J.}\ \bibnamefont {Williams}}, \bibinfo
  {author} {\bibfnamefont {I.~B.}\ \bibnamefont {Spielman}}, \ and\ \bibinfo
  {author} {\bibfnamefont {I.~I.}\ \bibnamefont {Satija}},\ }\href {\doibase
  10.1103/PhysRevA.84.063629} {\bibfield  {journal} {\bibinfo  {journal} {Phys.
  Rev. A}\ }\textbf {\bibinfo {volume} {84}},\ \bibinfo {pages} {063629}
  (\bibinfo {year} {2011})}\BibitemShut {NoStop}%
\bibitem [{\citenamefont {Alba}\ \emph {et~al.}(2011)\citenamefont {Alba},
  \citenamefont {Fernandez-Gonzalvo}, \citenamefont {Mur-Petit}, \citenamefont
  {Pachos}, \citenamefont {Garc{\'\i}a-Ripoll} \emph
  {et~al.}}]{alba2011seeing}%
  \BibitemOpen
  \bibfield  {author} {\bibinfo {author} {\bibfnamefont {E.}~\bibnamefont
  {Alba}}, \bibinfo {author} {\bibfnamefont {X.}~\bibnamefont
  {Fernandez-Gonzalvo}}, \bibinfo {author} {\bibfnamefont {J.}~\bibnamefont
  {Mur-Petit}}, \bibinfo {author} {\bibfnamefont {J.}~\bibnamefont {Pachos}},
  \bibinfo {author} {\bibfnamefont {J.~J.}\ \bibnamefont {Garc{\'\i}a-Ripoll}},
   \emph {et~al.},\ }\href@noop {} {\bibfield  {journal} {\bibinfo  {journal}
  {Phys. Rev. Lett.}\ }\textbf {\bibinfo {volume} {107}},\ \bibinfo {pages}
  {235301} (\bibinfo {year} {2011})}\BibitemShut {NoStop}%
\bibitem [{\citenamefont {Price}\ and\ \citenamefont
  {Cooper}(2012)}]{price2012mapping}%
  \BibitemOpen
  \bibfield  {author} {\bibinfo {author} {\bibfnamefont {H.}~\bibnamefont
  {Price}}\ and\ \bibinfo {author} {\bibfnamefont {N.}~\bibnamefont {Cooper}},\
  }\href@noop {} {\bibfield  {journal} {\bibinfo  {journal} {Phys. Rev. A}\
  }\textbf {\bibinfo {volume} {85}},\ \bibinfo {pages} {033620} (\bibinfo
  {year} {2012})}\BibitemShut {NoStop}%
\bibitem [{\citenamefont {Liu}\ \emph {et~al.}(2013)\citenamefont {Liu},
  \citenamefont {Law}, \citenamefont {Ng},\ and\ \citenamefont
  {Lee}}]{liu2013detecting}%
  \BibitemOpen
  \bibfield  {author} {\bibinfo {author} {\bibfnamefont {X.-J.}\ \bibnamefont
  {Liu}}, \bibinfo {author} {\bibfnamefont {K.}~\bibnamefont {Law}}, \bibinfo
  {author} {\bibfnamefont {T.}~\bibnamefont {Ng}}, \ and\ \bibinfo {author}
  {\bibfnamefont {P.~A.}\ \bibnamefont {Lee}},\ }\href@noop {} {\bibfield
  {journal} {\bibinfo  {journal} {Phys. Rev. Lett.}\ }\textbf {\bibinfo
  {volume} {111}},\ \bibinfo {pages} {120402} (\bibinfo {year}
  {2013})}\BibitemShut {NoStop}%
\bibitem [{\citenamefont {Abanin}\ \emph {et~al.}(2013)\citenamefont {Abanin},
  \citenamefont {Kitagawa}, \citenamefont {Bloch},\ and\ \citenamefont
  {Demler}}]{abanin2013interferometric}%
  \BibitemOpen
  \bibfield  {author} {\bibinfo {author} {\bibfnamefont {D.~A.}\ \bibnamefont
  {Abanin}}, \bibinfo {author} {\bibfnamefont {T.}~\bibnamefont {Kitagawa}},
  \bibinfo {author} {\bibfnamefont {I.}~\bibnamefont {Bloch}}, \ and\ \bibinfo
  {author} {\bibfnamefont {E.}~\bibnamefont {Demler}},\ }\href@noop {}
  {\bibfield  {journal} {\bibinfo  {journal} {Phys. Rev. Lett.}\ }\textbf
  {\bibinfo {volume} {110}},\ \bibinfo {pages} {165304} (\bibinfo {year}
  {2013})}\BibitemShut {NoStop}%
\bibitem [{\citenamefont {Atala}\ \emph {et~al.}(2013)\citenamefont {Atala},
  \citenamefont {Aidelsburger}, \citenamefont {Barreiro}, \citenamefont
  {Abanin}, \citenamefont {Kitagawa}, \citenamefont {Demler},\ and\
  \citenamefont {Bloch}}]{Atala2013direct}%
  \BibitemOpen
  \bibfield  {author} {\bibinfo {author} {\bibfnamefont {M.}~\bibnamefont
  {Atala}}, \bibinfo {author} {\bibfnamefont {M.}~\bibnamefont {Aidelsburger}},
  \bibinfo {author} {\bibfnamefont {J.~T.}\ \bibnamefont {Barreiro}}, \bibinfo
  {author} {\bibfnamefont {D.}~\bibnamefont {Abanin}}, \bibinfo {author}
  {\bibfnamefont {T.}~\bibnamefont {Kitagawa}}, \bibinfo {author}
  {\bibfnamefont {E.}~\bibnamefont {Demler}}, \ and\ \bibinfo {author}
  {\bibfnamefont {I.}~\bibnamefont {Bloch}},\ }\href@noop {} {\bibfield
  {journal} {\bibinfo  {journal} {Nat. Phys.}\ }\textbf {\bibinfo {volume}
  {9}},\ \bibinfo {pages} {795} (\bibinfo {year} {2013})}\BibitemShut {NoStop}%
 \bibitem [{\citenamefont {Goldman}\ \emph {et~al.}(2013)\citenamefont
  {Goldman}, \citenamefont {Anisimovas}, \citenamefont {Gerbier}, \citenamefont
  {{\"O}hberg}, \citenamefont {Spielman},\ and\ \citenamefont
  {Juzeli{\=u}nas}}]{Goldman2013Measuring}%
  \BibitemOpen
  \bibfield  {author} {\bibinfo {author} {\bibfnamefont {N.}~\bibnamefont
  {Goldman}}, \bibinfo {author} {\bibfnamefont {E.}~\bibnamefont {Anisimovas}},
  \bibinfo {author} {\bibfnamefont {F.}~\bibnamefont {Gerbier}}, \bibinfo
  {author} {\bibfnamefont {P.}~\bibnamefont {{\"O}hberg}}, \bibinfo {author}
  {\bibfnamefont {I.~B.}\ \bibnamefont {Spielman}}, \ and\ \bibinfo {author}
  {\bibfnamefont {G.}~\bibnamefont {Juzeli{\=u}nas}},\ }\href
  {http://stacks.iop.org/1367-2630/15/i=1/a=013025} {\bibfield  {journal}
  {\bibinfo  {journal} {New J. Phys.}\ }\textbf {\bibinfo {volume} {15}},\
  \bibinfo {pages} {013025} (\bibinfo {year} {2013})}\BibitemShut {NoStop}%
\bibitem{Pachos2013}
J. K. Pachos, E. Alba, V. Lahtinen, and J. J. Garcia-Ripoll, Phys. Rev. A \textbf{88}, 013622 (2013).
\bibitem{Lisle2014}
 J. de Lisle, S. De, E. Alba, A. Bullivant, J. J. Garcia-Ripoll, V. Lahtinen, and J. K. Pachos, arXiv:1402.3222v2.

\bibitem [{\citenamefont {Ketterle}\ and\ \citenamefont
  {Zwierlein}(2008)}]{ketterle2008making}%
  \BibitemOpen
  \bibfield  {author} {\bibinfo {author} {\bibfnamefont {W.}~\bibnamefont
  {Ketterle}}\ and\ \bibinfo {author} {\bibfnamefont {M.}~\bibnamefont
  {Zwierlein}},\ }\href@noop {} {\bibfield  {journal} {\bibinfo  {journal}
  {Riv. Nuovo Cimento}\ }\textbf {\bibinfo {volume} {31}},\ \bibinfo {pages}
  {247} (\bibinfo {year} {2008})}\BibitemShut {NoStop}%
\bibitem [{\citenamefont {Duan}(2006)}]{duan2006detecting}%
  \BibitemOpen
  \bibfield  {author} {\bibinfo {author} {\bibfnamefont {L.-M.}\ \bibnamefont
  {Duan}},\ }\href@noop {} {\bibfield  {journal} {\bibinfo  {journal} {Phys.
  Rev. Lett.}\ }\textbf {\bibinfo {volume} {96}},\ \bibinfo {pages} {103201}
  (\bibinfo {year} {2006})}\BibitemShut {NoStop}%
\bibitem [{\citenamefont {Fukui}\ \emph {et~al.}(2005)\citenamefont {Fukui},
  \citenamefont {Hatsugai},\ and\ \citenamefont {Suzuki}}]{fukui2005chern}%
  \BibitemOpen
  \bibfield  {author} {\bibinfo {author} {\bibfnamefont {T.}~\bibnamefont
  {Fukui}}, \bibinfo {author} {\bibfnamefont {Y.}~\bibnamefont {Hatsugai}}, \
  and\ \bibinfo {author} {\bibfnamefont {H.}~\bibnamefont {Suzuki}},\
  }\href@noop {} {\bibfield  {journal} {\bibinfo  {journal} {J. Phys. Soc.
  Jpn.}\ }\textbf {\bibinfo {volume} {74}},\ \bibinfo {pages} {1674} (\bibinfo
  {year} {2005})}\BibitemShut {NoStop}%
\bibitem [{\citenamefont {Haldane}(1988)}]{haldane1988model}%
  \BibitemOpen
  \bibfield  {author} {\bibinfo {author} {\bibfnamefont {F.}~\bibnamefont
  {Haldane}},\ }\href@noop {} {\bibfield  {journal} {\bibinfo  {journal} {Phys.
  Rev. Lett.}\ }\textbf {\bibinfo {volume} {61}},\ \bibinfo {pages} {2015}
  (\bibinfo {year} {1988})}\BibitemShut {NoStop}%
\bibitem [{\citenamefont {Chang}\ \emph {et~al.}(2013)\citenamefont {Chang},
  \citenamefont {Zhang}, \citenamefont {Feng}, \citenamefont {Shen},
  \citenamefont {Zhang}, \citenamefont {Guo}, \citenamefont {Li}, \citenamefont
  {Ou}, \citenamefont {Wei}, \citenamefont {Wang} \emph
  {et~al.}}]{chang2013experimental}%
  \BibitemOpen
  \bibfield  {author} {\bibinfo {author} {\bibfnamefont {C.-Z.}\ \bibnamefont
  {Chang}}, \bibinfo {author} {\bibfnamefont {J.}~\bibnamefont {Zhang}},
  \bibinfo {author} {\bibfnamefont {X.}~\bibnamefont {Feng}}, \bibinfo {author}
  {\bibfnamefont {J.}~\bibnamefont {Shen}}, \bibinfo {author} {\bibfnamefont
  {Z.}~\bibnamefont {Zhang}}, \bibinfo {author} {\bibfnamefont
  {M.}~\bibnamefont {Guo}}, \bibinfo {author} {\bibfnamefont {K.}~\bibnamefont
  {Li}}, \bibinfo {author} {\bibfnamefont {Y.}~\bibnamefont {Ou}}, \bibinfo
  {author} {\bibfnamefont {P.}~\bibnamefont {Wei}}, \bibinfo {author}
  {\bibfnamefont {L.-L.}\ \bibnamefont {Wang}},  \emph {et~al.},\ }\href@noop
  {} {\bibfield  {journal} {\bibinfo  {journal} {Science}\ }\textbf {\bibinfo
  {volume} {340}},\ \bibinfo {pages} {167} (\bibinfo {year}
  {2013})}\BibitemShut {NoStop}%
\bibitem [{\citenamefont {Liu}\ \emph {et~al.}(2014)\citenamefont {Liu},
  \citenamefont {Law},\ and\ \citenamefont {Ng}}]{Liu2014Realization}%
  \BibitemOpen
  \bibfield  {author} {\bibinfo {author} {\bibfnamefont {X.-J.}\ \bibnamefont
  {Liu}}, \bibinfo {author} {\bibfnamefont {K.~T.}\ \bibnamefont {Law}}, \ and\
  \bibinfo {author} {\bibfnamefont {T.~K.}\ \bibnamefont {Ng}},\ }\href@noop {}
  {\bibfield  {journal} {\bibinfo  {journal} {Phys. Rev. Lett.}\ }\textbf
  {\bibinfo {volume} {112}},\ \bibinfo {pages} {086401} (\bibinfo {year}
  {2014})}\BibitemShut {NoStop}%
\bibitem [{sup()}]{supplement}%
  \BibitemOpen
  \href@noop {} {}\bibinfo {note} {See Supplemental Material at [URL will be
  inserted by publisher] for more details on the calculations of the momentum density distributions and more plots of the numerical simulations.}\BibitemShut {NoStop}%
\bibitem [{\citenamefont {Neupert}\ \emph {et~al.}(2012)\citenamefont
  {Neupert}, \citenamefont {Santos}, \citenamefont {Ryu}, \citenamefont
  {Chamon},\ and\ \citenamefont {Mudry}}]{neupert2012noncommutative}%
  \BibitemOpen
  \bibfield  {author} {\bibinfo {author} {\bibfnamefont {T.}~\bibnamefont
  {Neupert}}, \bibinfo {author} {\bibfnamefont {L.}~\bibnamefont {Santos}},
  \bibinfo {author} {\bibfnamefont {S.}~\bibnamefont {Ryu}}, \bibinfo {author}
  {\bibfnamefont {C.}~\bibnamefont {Chamon}}, \ and\ \bibinfo {author}
  {\bibfnamefont {C.}~\bibnamefont {Mudry}},\ }\href@noop {} {\bibfield
  {journal} {\bibinfo  {journal} {Phys. Rev. B}\ }\textbf {\bibinfo {volume}
  {86}},\ \bibinfo {pages} {035125} (\bibinfo {year} {2012})}\BibitemShut
  {NoStop}%
\bibitem [{\citenamefont {Wang}\ \emph {et~al.}(2014)\citenamefont {Wang},
  \citenamefont {Deng},\ and\ \citenamefont {Duan}}]{Wang2014probe}%
  \BibitemOpen
  \bibfield  {author} {\bibinfo {author} {\bibfnamefont {S.-T.}\ \bibnamefont
  {Wang}}, \bibinfo {author} {\bibfnamefont {D.-L.}\ \bibnamefont {Deng}}, \
  and\ \bibinfo {author} {\bibfnamefont {L.-M.}\ \bibnamefont {Duan}},\ }\href
  {\doibase 10.1103/PhysRevLett.113.033002}
   {\bibfield  {journal} {\bibinfo
  {journal} {Phys. Rev. Lett.}\ }\textbf {\bibinfo {volume} {113}},\ \bibinfo
  {pages} {033002} (\bibinfo {year} {2014})}\BibitemShut {NoStop}%
  \end{thebibliography}

\begin{thebibliography}{{citenamefont{Alba et~al.}(2011)citenamefont{Alba,   Fernandez-Gonzalvo, Mur-Petit, Pachos, Garc{\'\i}a-Ripoll   et~al.}}}
\bibitem[{citenamefont{Alba et~al.}(2011)citenamefont{Alba,   Fernandez-Gonzalvo, Mur-Petit, Pachos, Garc{\'{}\i{}}a-Ripoll   et~al.}}]{alba2011seeing}
\bibinfo{author}{\bibfnamefont{E.}~\bibnamefont{Alba}},
\bibinfo{author}{\bibfnamefont{X.}~\bibnamefont{Fernandez-Gonzalvo}},
\bibinfo{author}{\bibfnamefont{J.}~\bibnamefont{Mur-Petit}},
\bibinfo{author}{\bibfnamefont{J.}~\bibnamefont{Pachos}},
\bibinfo{author}{\bibfnamefont{J.~J.} \bibnamefont{Garc{í}a-Ripoll}},
\bibnamefont{et~al.}, \bibinfo{journal}{Phys. Rev. Lett.} \textbf{\bibinfo{volume}{107}},
\bibinfo{pages}{235301} (\bibinfo{year}{2011}).
\bibitem [{\citenamefont {Wang}\ \emph {et~al.}(2014)\citenamefont {Wang},
  \citenamefont {Deng},\ and\ \citenamefont {Duan}}]{Wang2014probe}%
  \BibitemOpen
  \bibfield  {author} {\bibinfo {author} {\bibfnamefont {S.-T.}\ \bibnamefont
  {Wang}}, \bibinfo {author} {\bibfnamefont {D.-L.}\ \bibnamefont {Deng}}, \
  and\ \bibinfo {author} {\bibfnamefont {L.-M.}\ \bibnamefont {Duan}},\ }
   {\bibfield  {journal} {\bibinfo
  {journal} {Phys. Rev. Lett.}\ }\textbf {\bibinfo {volume} {113}},\ \bibinfo
  {pages} {033002} (\bibinfo {year} {2014})}\BibitemShut {NoStop}%
\bibitem[{citenamefont{Neupert et~al.}(2012)citenamefont{Neupert, Santos,   Ryu, Chamon, and Mudry}}]{neupert2012noncommutative}
\bibinfo{author}{\bibfnamefont{T.}~\bibnamefont{Neupert}},
\bibinfo{author}{\bibfnamefont{L.}~\bibnamefont{Santos}},
\bibinfo{author}{\bibfnamefont{S.}~\bibnamefont{Ryu}}, \bibinfo{author}{\bibfnamefont{C.}~\bibnamefont{Chamon}},
\bibnamefont{and} \bibinfo{author}{\bibfnamefont{C.}~\bibnamefont{Mudry}},
\bibinfo{journal}{Phys. Rev. B} \textbf{\bibinfo{volume}{86}},
\bibinfo{pages}{035125} (\bibinfo{year}{2012}).\end{thebibliography}

\expandafter\ifx\csname natexlab\endcsname\relax\global\long\def\natexlab#1{#1}
\fi \expandafter\ifx\csname bibnamefont\endcsname\relax \global\long\def\bibnamefont#1{#1}
\fi \expandafter\ifx\csname bibfnamefont\endcsname\relax \global\long\def\bibfnamefont#1{#1}
\fi \expandafter\ifx\csname citenamefont\endcsname\relax \global\long\def\citenamefont#1{#1}
\fi \expandafter\ifx\csname url\endcsname\relax \global\long\def\url#1{\texttt{#1}}
\fi \expandafter\ifx\csname urlprefix\endcsname\relax\global\long\def\urlprefix{URL }
\fi \providecommand{\bibinfo}[2]{#2} \providecommand{\eprint}[2][]{\url{#2}}

\end{document}